\UseRawInputEncoding

\documentclass[aps,prl,amsmath,amssymb,floatfix,reprint,citeautoscript,noeprint,superscriptaddress,twocolumn]{revtex4-2}
\usepackage{lineno}

\usepackage{dsfont}
\usepackage{lipsum} 
\usepackage{bibentry}
\usepackage[english]{babel}
\usepackage[dvipsnames]{xcolor}
\usepackage{graphicx}
\usepackage[caption=false]{subfig} 
\usepackage{amsmath,amssymb,bm}
\usepackage[version=3]{mhchem}
\usepackage{verbatim}
\usepackage{multirow}
\usepackage{dcolumn}
\usepackage{float}
\usepackage{nicefrac}
\usepackage{siunitx}
\usepackage{booktabs}
\usepackage{chemformula}
\usepackage{wrapfig}
\usepackage{enumitem}  
\usepackage{transparent}
\usepackage[colorlinks,allcolors=black,citecolor=blue,urlcolor=blue]{hyperref}
\usepackage{tcolorbox}
\usepackage{relsize}
\hypersetup{
    citecolor=Turquoise,
    colorlinks=true,
    linkcolor=black,    
    urlcolor=Turquoise,
    }
\usepackage{titlesec}

\titleformat{\section}
  {\normalfont\normalsize\bfseries} % Normal font, same size as text, bold
  {} % No section number
  {0pt} % No spacing
  {} % No additional formatting
  [\vspace{-5pt}] 
    
\DeclareSIUnit[number-unit-product = {\,}]{\amu}{amu}
\DeclareSIUnit[number-unit-product = {\,}]{\kJmol}{\kilo\joule\per\mol}
\DeclareSIUnit[number-unit-product = {\,}]{\Nsm}{\newton\second\per\meter\cubed}
\DeclareSIUnit[number-unit-product = {\,}]{\THz}{\tera\hertz}
\DeclareSIUnit[number-unit-product = {\,}]{\meV}{\milli\electronvolt}
\DeclareSIUnit[number-unit-product = {\,}]{\cal}{cal}

\newcommand{\mbf}[1]{\boldsymbol{\mathit{#1}}}
\newcommand{\mrm}[1]{\mathrm{#1}}
\newcommand{\mcl}[1]{\mathcal{#1}}
\newcommand{\tcr}[1]{\textcolor{black}{#1}}

\newcommand{\abinitio}{\emph{ab initio }}

\usepackage{sansmathfonts}
\usepackage[justification=centerlast, font={sf}, labelfont={bf}]{caption}
\begin{document}

\title{A unified machine-learning framework for \emph{ab initio} multiscale modeling of liquids}

\author{Anna T. Bui}
\affiliation{Yusuf Hamied Department of Chemistry, University of
  Cambridge, Lensfield Road, Cambridge, CB2 1EW, United Kingdom}
\affiliation{Department of Chemistry, Durham University, South Road,
  Durham, DH1 3LE, United Kingdom}

\author{Stephen J. Cox}
\email{stephen.j.cox@durham.ac.uk}
\affiliation{Department of Chemistry, Durham University, South Road,
  Durham, DH1 3LE, United Kingdom}

\date{\today}

\color{darkgray}
%\pagecolor{lightgray}

\begin{abstract}
  \textbf{Abstract:} Understanding and predicting the behavior of
  liquid matter across length scales---using only the microscopic
  interactions encoded in the Schr\"odinger equation---remains a
  central challenge in the physical sciences. Achieving this goal
  requires not only an accurate and efficient description of
  intermolecular forces but also a consistent framework that bridges
  the micro-, meso-, and macroscales. Here, by combining
  machine-learned interatomic potentials (MLIPs) with neural classical
  density functional theory (cDFT), we present such a
  framework. MLIPs trained on quantum-mechanical energies and forces
  are used to generate inhomogeneous density profiles, which
  then serve as the training data for neural cDFT. The resulting
  \emph{ab initio} neural cDFT is more computationally efficient than
  molecular simulations and provides a conceptually transparent route
  to the thermodynamics of both homogeneous and planar
  inhomogeneous systems. We demonstrate the approach for both water
  and carbon dioxide using several exchange--correlation
  functionals. Beyond accurately reproducing---at the level of
    the underlying approximate electronic structure---bulk equations
  of state and liquid--vapor phase diagrams, \emph{ab initio} neural
  cDFT predicts, from first principles, how confinement modifies
  liquid--vapor coexistence in water. It also captures complex
  behavior in supercritical carbon dioxide such as the Fisher--Widom
  and Widom lines. While current applications are limited to bulk
    fluids and planar geometries, this approach establishes a general
    first-principles route to multiscale modeling of fluids by
    unifying two independently developed machine-learning paradigms.
    This work represents an important step toward generalizing cDFT
    beyond simple empirical potentials to chemically complex systems.
\end{abstract}

\maketitle

Liquids play a pivotal role across biology, energy storage, catalysis,
and environmental science. Their influence ranges from processes at
the molecular scale \cite{Geissler2001, Zhang2025, Zhang2024}, where
the quantum nature of interatomic interactions is important, to the
nano-, meso-, and macroscopic redistribution of the fluid
\cite{Perkin2010, Zhang2026, Liu2017}, such as near phase
transitions. Put simply, comprehensively understanding the behavior of
liquids is an inherently multiscale problem. A central goal in
chemical physics is to predict such multiscale phenomena from first
principles, relying solely on knowledge of the underlying microscopic
interactions encoded in the Schr\"{o}dinger equation
\cite{Radhakrishnan2021}.  Any genuine multiscale modeling approach
needs to faithfully describe both a system's intermolecular
interactions, and its collective behavior.

Rooted in the majority of existing multiscale frameworks is the
notion of different ``levels of theory.'' For example, in the case of
QM/MM \cite{Field1990} a region where interactions are accurately
described by quantum mechanics (QM) is embedded within an environment
(e.g., solvent) described by a computationally efficient molecular
mechanics (MM) model. Alternatively, the results of accurate \emph{ab
  initio} calculations can be used to parameterize a coarser
description of the system, e.g., by passing reaction energies into a
kinetic model \cite{Reuter2006}. While such strategies have proven
powerful in contexts such as biochemistry \cite{Karplus2002} and
heterogeneous catalysis \cite{Bruix2019}, how to best combine
different levels of theory is not always apparent. In this article, we
demonstrate a simple multiscale modeling strategy for liquids that
simultaneously describes micro-, meso-, and macroscale phenomena on an
equal footing, entirely from first principles.

The multiscale framework that we present unifies two areas of physical
science in which machine learning (ML) is playing a transformative
role. The first of these is machine-learned interatomic potentials
(MLIPs). By providing relatively inexpensive surrogate models of a
system's potential energy surface that would otherwise be obtained
from costly electronic structure calculations, MLIPs are vastly
increasing the efficiency of molecular simulations of condensed
phases. The second is classical density functional theory (cDFT), an
exact statistical mechanical framework for inhomogeneous fluids that,
in principle, provides mesoscale insight while retaining microscopic
resolution \cite{HansenMcDonaldBook, Evans1979}.

The development of MLIPs is an avenue of research actively pursued by
many research groups \cite{Behler2007, Zhang2018, Batatia2022,
  wood2025, rhodes2025, Mazitov2025}, and is sufficiently advanced
that predicting microscopic structure, dynamics, and bulk
thermodynamics of fluids has essentially become routine
\cite{Schran2021, Cheng2019, Wohlfahrt2020, Zhang2022natcom, Mathur2023, Magdau2023}.
Yet, while direct simulation using MLIPs to compute emergent
properties such as phase diagrams \cite{Zong2020, Reinhardt2021,
  Kapil2022, Menon2024}, interfacial free energies \cite{Zhang2025},
and adsorption equilibria \cite{Goeminne2025} is possible, doing so
comes with a significant computational burden. Moreover, such
simulations are often sensitive to simulation size, slow dynamics
associated with rare events, and largely limited to describing closed
systems; standard molecular dynamics (MD) simulations are typically
performed with a fixed number of molecules. For systems in equilibrium
with a reservoir, as is often the case for adsorption and confinement
phenomena, a framework that naturally lends itself to open systems can
greatly simplify analysis.

\begin{figure*}[t]
  \includegraphics[width=0.9\linewidth]{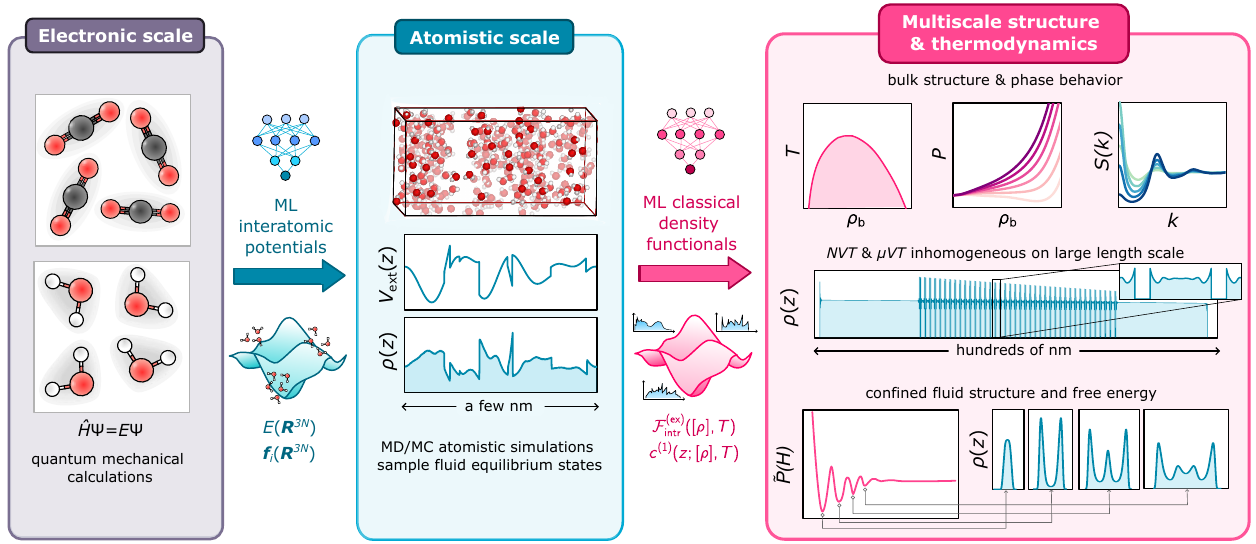}
  \caption{\textbf{Overview of \abinitio neural cDFT.}  Energies and
    forces from small-scale electronic structure calculations are used
    to train an MLIP that represents the potential energy surface,
    enabling efficient sampling of atomic configurations on nanometer
    length scales. Equilibrium density profiles obtained from
    molecular simulations with the MLIP under inhomogeneous external
    potentials then form the training set for neural cDFT. The
    resulting \emph{ab initio} neural cDFT can be used to obtain bulk
    thermophysical properties, liquid--vapor phase equilibria, and to
    investigate inhomogeneous systems both on large length scales, and
    under nanoconfinement.}
\label{fig1}
\end{figure*}

To overcome the intrinsic limitations of molecular simulation, we will
also leverage recent advances in cDFT, in which, at temperature $T$, both a fluid's equilibrium one-body density $\rho(\mbf{r})$
and thermodynamics are obtained by minimizing the system's grand
potential functional
\begin{eqnarray}
  \varOmega_{V}([\rho],T) & = \mcl{F}^{\rm (id)}_{\rm intr}([\rho],T) 
                               + \mcl{F}^{\mrm{(ex)}}_{\mrm{intr}}([\rho],T) \nonumber \\
                             & + \int\!\!\mrm{d}\mbf{r}\, \rho(\mbf{r}) \left( V_{\rm ext}(\mbf{r})- \mu \right),
\end{eqnarray}
in a single, self-consistent calculation \cite{HansenMcDonaldBook,
  Evans1979}. Inhomogeneity in $\rho$ may arise from the external
potential $V_{\rm ext}$, or from the formation of interfaces between
phases at coexistence. The intrinsic Helmholtz free energy functional,
which is independent of $V_{\rm ext}$ and the chemical potential
$\mu$, comprises an ideal part $\mathcal{F}^{\rm (id)}_{\rm intr}$
that is known exactly, and an excess part
$\mathcal{F}^{\rm (ex)}_{\rm intr}$ that arises from intermolecular
interactions.

In direct analogy to the exchange--correlation (xc) functional in
electronic structure \cite{Kohn1999}, the use of cDFT as a practical
tool relies upon accurate approximations to
$\mathcal{F}^{\rm (ex)}_{\rm intr}$.  Note that, in contrast to the
exact xc functional in electronic structure,
$\mathcal{F}^{\rm (ex)}_{\rm intr}$ is not universal; it depends upon
the underlying intermolecular potential and $T$. For a given
intermolecular potential, however, $\mathcal{F}^{\rm (ex)}_{\rm intr}$
is unique.  While accurate forms of
$\mathcal{F}^{\rm (ex)}_{\rm intr}$ have been available for hard
sphere fluids for decades \cite{Rosenfeld1989, Roth2002, Roth2010},
going beyond simple liquids, where hard spheres can act as a suitable
reference, has remained a formidable challenge \cite{Tang2004,
  Jeanmairet2013, Archer2017, Jeanmairet2026}.

Recent works have shown that ML can be used to learn highly accurate
representations of $\mcl{F}^{\mathrm{(ex)}}_{\mrm{intr}}$ from
molecular simulation data \cite{Simon2025, Dijkman2025, Yang2025,
  Simon2026}.  Specifically, Samm\"{u}ller \emph{et al.}, in a
development dubbed ``neural cDFT," have shown how a neural network can
be trained to represent the one-body direct correlation functional
\cite{Sammuller2023},
\begin{equation}
  \label{eqn:c1-def}
  c^{(1)}(\mbf{r};[\rho], T) = -\frac{\delta\beta \mcl{F}^{\rm (ex)}_{\rm intr}([\rho], T)}{\delta\rho(\mbf{r})},
\end{equation}
where $\beta =1/k_{\rm B}T$ with $k_{\rm B}$ as the Boltzmann
constant. Neural cDFT has been successfully applied to a range of
  model fluids, including hard spheres, Lennard--Jones (both single-
  and two-component), primitive electrolytes, and polar fluids
  \cite{Sammuller2025, Robitschko2025, zhou2026, Bui2025prl,
    bui2025dielectro}, where it has been shown to describe complex
  phenomena such as liquid--vapor coexistence, liquid--liquid phase
  separation, azeotropy, and electromechanical phenomena. All of these
  previous studies, however, have relied exclusively on empirical
interatomic potentials, leaving cDFT's potential to predict mesoscopic
behavior directly from a first-principles Hamiltonian untapped.

Here, we fill this gap by presenting an integration of MLIPs
  trained on quantum-mechanical data into neural cDFT, yielding a
  genuinely \emph{ab initio} cDFT for molecular fluids. We
  demonstrate this framework for two liquids of broad importance:
water and carbon dioxide. After validating the resulting \emph{ab
  initio} neural cDFT against molecular simulations, we use it to
straightforwardly investigate phenomena that would be extremely
challenging to obtain with traditional computational
approaches. Specifically, we investigate the influence of confinement
on water's liquid--vapor phase behavior, with a clear thermodynamics
prescribed by the grand canonical ensemble. We also compute the fluid
phase diagram of carbon dioxide, including the structural crossovers
encoded by the Fisher--Widom and Widom lines. A schematic overview of
our approach is given in Fig.~\ref{fig1}.

\section*{Constructing \abinitio neural cDFT}

Rather than use MLIPs to study the behavior of fluids directly, we
instead use their predicted forces to generate equilibrium planar
inhomogeneous density profiles $\rho_{\rm eq}(z)$ from relatively
small simulations. Specifically, for a given $T$, total number of
molecules, and $V_{\rm ext}(z)$ (each chosen randomly), we obtain
$\rho_{\mrm{eq}}(z)$ from an MD simulation. The corresponding one-body
direct correlation function is given by the Euler--Lagrange equation
that results from the variational principle of cDFT
\cite{HansenMcDonaldBook}
\begin{equation}
  \label{eqn:EL}
  c^{(1)}(z;[\rho_{\mrm{eq}}],T) = \ln\big(\zeta^{-1}\Lambda^3\rho_{\mrm{eq}}(z)\big) + \beta\big(V_{\rm ext}(z)-\mu\big),
\end{equation}
where $\Lambda$ is the thermal de Broglie wavelength, and $\zeta$ is
an intramolecular partition function; this is necessary as $\mu$
couples to the number of molecules rather than atoms
\cite{bui2025hyperlmft}. For the remainder of the article, since
  we only consider the system at equilibrium, we hereinafter drop the
  ``eq'' subscript, such that $\rho$ refers to the equilibrium density
  profile.

In the original implementations of neural cDFT, many grand canonical
Monte Carlo (GCMC) simulations, i.e., those with different, but known,
$V_{\rm ext}$ and $\mu$, were used to generate a training set by
obtaining $c^{(1)}(z)$ directly from Eq.~\ref{eqn:EL}. For each
discrete value of $c^{\rm (1)}(z)$, treated as the target in the
learning procedure, a finite window of neighboring points in the
corresponding $\rho(z)$ is passed through a deep neural network to
learn the functional dependence on density \cite{Sammuller2023}. The
temperature dependence can also be established by passing $T$ to the
neural network \cite{Sammuller2025}. For the MLIPs we use here,
however, GCMC is impractical; this is due to both a lack of efficient
software packages and the fact that insertion moves in GCMC can lead
to configurations far outside the MLIP training set
\cite{Goeminne2023}. We therefore adopt a recent development whereby
training data is obtained from canonical MD simulations
\cite{sammuller2026}, in which both $c^{(1)}$ and the set of chemical
potentials of the training simulations are learned by the neural
network; this is achieved by treating Eq.~\ref{eqn:EL} itself as the
loss function (see \emph{Methods}). Thus, $\mu$ is treated as a
latent variable in the machine learning problem.

The neural cDFT procedure that we describe above is limited to
  learning the one-body direct correlation functional of fluids with
  planar inhomogeneous geometries, though we stress that the systems
  remain three dimensional. This limitation is, in part, due to the
  underlying machine learning architecture, though we note recent
  efforts to extend neural cDFT to resolving the fluid density in two
  dimensions \cite{Glitsch2025}. Moreover, even if we were to attempt
  to adopt the approach described in Ref.~\onlinecite{Glitsch2025} to
  full three-dimensional resolution, obtaining reliable training data
  would currently prove too onerous for the MLIPs we use in this
  study. It is important to stress, however, that functions of the
  bulk fluid with radial symmetry, such as the radial two-body direct
  correlation function, can be obtained by automatic differentiation
  of $c^{(1)}(z)$ from neural cDFT, followed by radial projection
  \cite{Sammuller2023}.

As $\mathcal{F}^{(\rm ex)}_{\rm intr}$ is not a universal functional,
for each fluid and xc functional we must learn $c^{(1)}$
separately. For water, we therefore learn $c^{(1)}$ corresponding to
the SCAN \cite{Sun2015} and RPBE-D3 \cite{Hammer1999,Grimme2010}
functionals, as MLIPs for these have previously been successfully
applied to liquid--vapor coexistence. For the purposes of comparison,
we also present results for TIP4P/2005 \cite{Abascal2005}, a popular
empirical potential for studying liquid water. In the case of carbon
dioxide, we obtain $c^{(1)}$ for the PBE-D3 \cite{Perdew1996}, BLYP-D3
\cite{Becke1988}, and SCAN-rVV10 \cite{Peng2016} functionals, along
with the well-established TraPPE empirical potential
\cite{Potoff2001}. Although rooted in quantum mechanics, all xc
  functionals we consider are necessarily approximate representations
  of the true microscopic Hamiltonian. Our selection is guided
  primarily by the availability of extensive reference data---such as
  bulk thermodynamic properties and liquid--vapor coexistence---from
  prior MLIP-based simulation studies, which enables meaningful
  validation of our approach. While alternative electronic-structure
  methods could in principle provide more accurate benchmarks
  \cite{li2025assessment}, the \emph{ab initio} neural cDFT framework
  itself is agnostic to this choice, provided that suitable MLIPs can
  be constructed.

For each xc functional or empirical potential, between 500--1000
training simulations comprising 40--1024 molecules were
performed with different random $V_{\rm ext}$, and across a broad
range of $T$ (see \emph{Methods}). Representative $V_{\rm ext}$ and
$\rho(z)$ used during training are shown in
Fig.~\ref{fig2}A. Taking advantage of the canonical learning
  procedure, we also include $\rho(z)$ from simulations of
  liquid--vapor coexistence (comprising either 832 water molecules or
  1536 carbon dioxide molecules), in which $V_{\rm ext} = 0$. We
employed either DeepMD \cite{Zhang2018} or HD-NNP
\cite{Behler2007} for the underlying MLIP architecture (see
\emph{Methods}), however, we show in {Fig.~S1 that MACE
\cite{Batatia2022} also provides stable dynamics with the random forms
for $V_{\rm ext}$ that we use.

\section*{Validation of \emph{ab initio} prediction of structure and thermodynamics}

With a trained neural network representation of $c^{(1)}$, the
equilibrium structure for any given $T$, $V_{\rm ext}$, and $\mu$ is
readily obtained by self-consistently solving the Euler--Lagrange
equation (Eq.~\ref{eqn:EL}), rearranged here with $\rho$ as the object:
\begin{equation}
  \label{eqn:EL2}  
  \zeta^{-1}\Lambda^{3}\rho(z) = \exp\big[- \beta\big(V_{\mathrm{ext}}(z) - \mu\big) + c^{(1)}(z;[\rho],T)\big].
\end{equation}
Although $\rho$ represents a microscopic density field, it is an
average quantity; the microscopic degrees of freedom have been
integrated out. Accordingly, obtaining the equilibrium structure of
the fluid is significantly less computationally demanding than
performing the average explicitly with a molecular
simulation. Importantly, this efficiency is realized after a one-time
training stage: once the functional has been learned, it can be used
to perform thousands of structure and free energy calculations\tcr{, across many different thermodynamic state points,} 
at minimal additional
cost.  Furthermore, with neural cDFT,
$\mathcal{F}^{\rm (ex)}_{\rm intr}$ can be obtained by functional line
integration of $c^{(1)}$ (see \emph{Methods}), and subsequently used
to compute the grand potential $\Omega = \varOmega_{V}([\rho],T)$.  In
contrast, obtaining free energies from simulations is often an
involved and costly procedure comprising many intermediate steps
(e.g., thermodynamic integration).  To give a sense of the
computational gains, in what follows, each calculation with neural
cDFT that we present takes on the order of a minute on a standard
CPU/GPU; in about one hour wall-clock time on a single GPU (see
\emph{Methods}) we can predict $\rho(z)$---with full microscopic
resolution---over a $\sim\!200$\,nm extent (see Figs.~\ref{fig1}
  and~S2). Where direct comparison to molecular simulations is
possible, we also show that \emph{ab initio} neural cDFT is
accurate. In this section, we present a selection of results to
validate the approach, with a more extensive set of results shown in
Figs.~S3 and~S4.

\begin{figure*}[t]
  \includegraphics[width=0.9\linewidth]{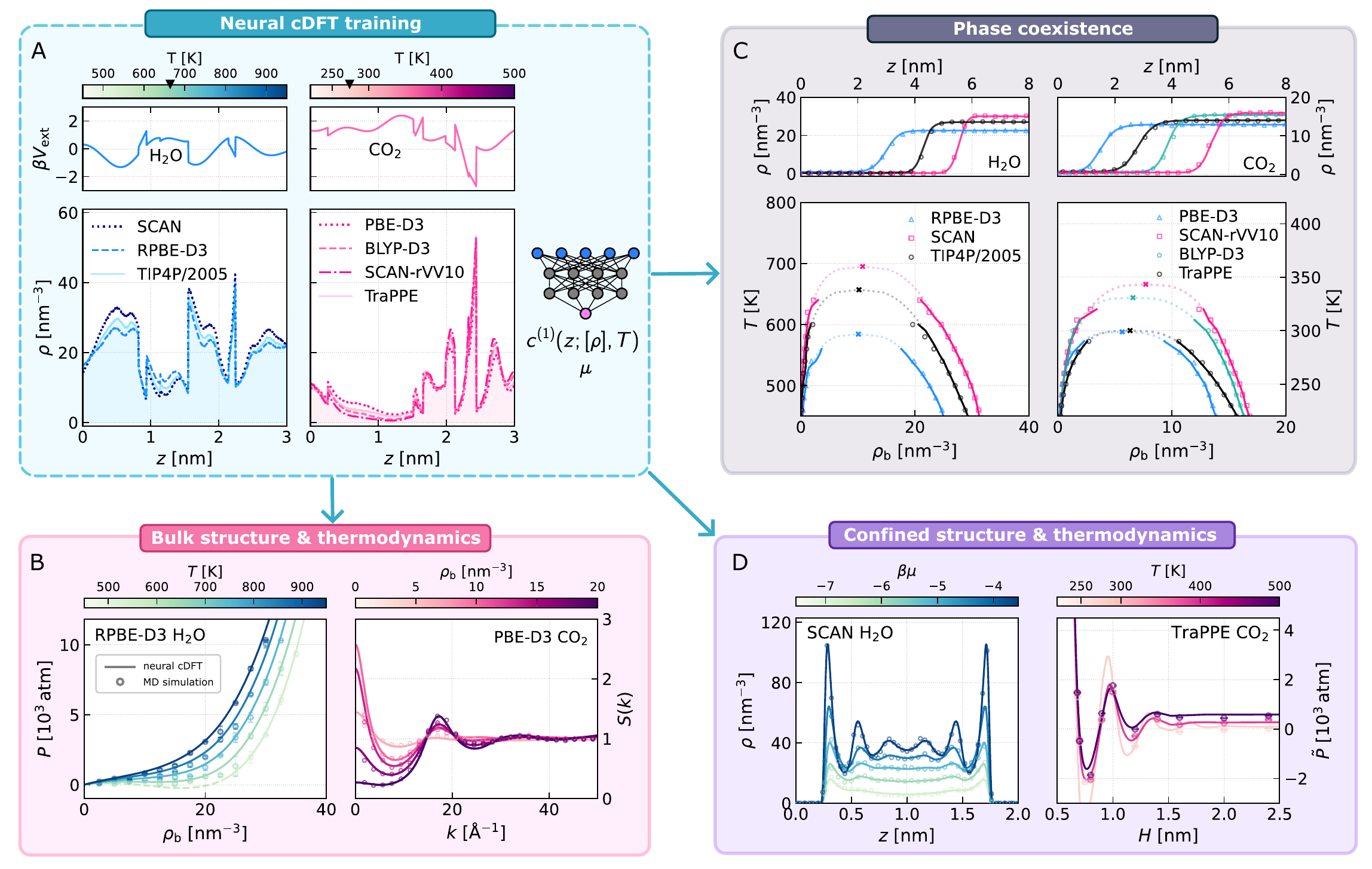}
  \caption{\textbf{Accurate and efficient description of liquids with
      \emph{ab initio} neural cDFT.} A: Typical random
    $V_{\rm ext}(z)$ and corresponding $\rho(z)$ used to train the
    neural cDFT, as obtained from molecular simulations. Results are
    shown for both water and carbon dioxide, with several interatomic
    potentials, as indicated in the legends. B, left: $P$ vs
    $\rho_{\rm b}$ for RPBE-D3 water along several isotherms. The
      van der Waals loop for the lowest temperature is indicated by
      the dashed line. B, right: $S(k)$ for PBE-D3 carbon dioxide
    for different $\rho_{\rm b}$ at $T=360$\,K. C, top: $\rho(z)$ at
    liquid--vapor coexistence for water at $T= 500\,$K (left) and
    carbon dioxide at $T=250\,$K (right), for different interatomic
    potentials. C, bottom: Liquid--vapor binodals for water (left) and
    carbon dioxide (right) with the different interatomic potentials,
    as indicated in the legend. D, left: $\rho(z)$ of SCAN water at
    different $\mu$ confined between graphene sheets. D, right:
    Effective pressure of TraPPE carbon dioxide at different $T$,
    confined between graphene sheets. In B--D, symbols show results
    from MD simulations and solid lines show results from \emph{ab
      initio} neural cDFT.}
\label{fig2}
\end{figure*}

\textbf{\emph{Bulk structure and thermodynamics}}. Beyond its
computational efficiency, an appealing feature of cDFT, especially in
the context of an \emph{ab initio} multiscale modeling framework, is
its direct access to thermodynamic properties. For example, for a
single-component bulk fluid with density $\rho_{\rm b}$, the pressure
can be obtained directly from $c^{(1)}$ and
$\mathcal{F}^{\rm (ex)}_{\rm intr}$ without the need for any
self-consistent calculation
\begin{equation}
  \label{eqn:pressure}
  P(\rho_{\rm b}, T) = k_{\rm B}T\rho_{\rm b}\left(1-c^{(1)}([\rho_{\rm b}], T)\right) - \frac{\mathcal{F}^{\rm (ex)}_{\rm intr}([\rho_{\rm b}], T)}{V}.
\end{equation}

In Fig.~\ref{fig2}B, we show $P$ against $\rho_{\rm b}$ for water at
different $T$, described by the RPBE-D3 xc functional. Results are
shown both for the neural cDFT using Eq.~\ref{eqn:pressure}, and
directly from molecular simulations. Not only does the neural cDFT
faithfully describe the simulation data, but it also exhibits a
  van der Waals loop at sufficiently low (subcritical) temperatures.
This observation is consistent with a previous neural cDFT study for a
Lennard--Jones fluid \cite{Sammuller2025}, though in contrast, we have
included coexistence simulations directly into the training
set. Nonetheless, we expect \emph{ab initio} neural cDFT---here
trained on data from relatively small molecular simulations with
MLIPs---to describe liquid--vapor phase equilibria. We explore this
point further below.

Despite only being trained on planar inhomogeneous density profiles,
the bulk structure of the fluid can also be obtained from neural
cDFT. Specifically, the bulk structure factor is provided by the
Ornstein--Zernike equation,
\begin{equation}
  \label{eqn:OZ}
  S(k; \rho_{\rm b}, T) =  \frac{1}{1- \rho_{\rm b}\hat{c}_{\rm r}^{(2)}(k; [\rho_{\rm b}], T)},
  % 1 + \rho_{\rm b}\hat{h}(k; \rho_{\rm b}, T) =
\end{equation}
where $\hat{c}^{(2)}_{\rm r}$ is the Fourier transform of the radial
two-body direct correlation function of the bulk fluid,
$c^{(2)}_{\rm r}$. As mentioned earlier, the latter is obtained
by automatic differentiation of $c^{(1)}$ from neural cDFT, followed
by radial projection \cite{Sammuller2023}.

Results for $S(k)$ are presented in Fig.~\ref{fig2}B for carbon
dioxide described by the PBE-D3 xc functional, at a
supercritical temperature $T=360\,$K. Again, we observe good agreement
between the \emph{ab initio} neural cDFT and results obtained directly
from molecular simulations. Note that $S(k\to 0)$ displays
non-monotonic behavior as $\rho_{\rm b}$ is varied. This observation
suggests that supercritical carbon dioxide exhibits a maximum in the
isothermal compressibility; we explore such supercritical behavior in
more detail later in the article.

\textbf{\emph{Liquid--vapor coexistence}}. We now consider
liquid--vapor phase equilibria. To construct the binodal in the
$\rho_{\rm b}$--$T$ plane, we follow a procedure analogous to a direct
coexistence simulation. Specifically, for a given $T$, we find a 
solution to Eq.~\ref{eqn:EL2} with $V_{\rm ext} = 0$, corresponding
to liquid--vapor coexistence, i.e., those with an interface between
the two phases; from the resulting density profile, we then simply
read the densities corresponding to the bulk liquid and vapor phases.

In order to stabilize interfacial solutions, we solve the
Euler--Lagrange equation subject to the constraint that the overall
density $\bar{\rho}_L = L^{-1}\int_0^L \!\mathrm{d}z\,\rho(z)$ is
fixed, where $L$ is the length of the system domain in
$z$. Whereas in typical cDFT calculations $\mu$ is specified as a
  control variable, here it acts as the Lagrange multiplier that
enforces the constraint. For the systems we investigate, this
procedure allows us to effectively mimic the canonical ensemble. This
approach is similar in spirit to that of Ref.~\cite{Sammuller2025},
which investigated liquid--vapor coexistence of the Lennard--Jones
fluid, though our approach to constraining $\bar{\rho}_L$ differs.

Results for both water and carbon dioxide are presented in
Fig.~\ref{fig2}C, with $L=20$\,nm. We also show results from direct
coexistence simulations. Overall, we observe excellent agreement
between the neural cDFT and the molecular
simulations. From the computed binodals, we estimate the
critical point from the empirical law of rectilinear diameters and
critical exponents \cite{rowlinson2002book}, yielding critical
temperature $T_{\rm c}$ and critical density $\rho_{\rm c}$.

For water, the neural cDFT predictions agree well with previous
simulation studies: TIP4P/2005 yields 
$T_{\mrm{c}}$ of 657\,K (cf. 640\,K from Vega \emph{et al.}
\cite{Vega2006}), RPBE-D3 gives 584\,K \cite{Wohlfahrt2020}, and SCAN
gives 695\,K \cite{Sanchez-Burgos2023}. For carbon dioxide, we obtain
300\,K for TraPPE, 299\,K for PBE-D3, 331\,K for BLYP-D3, and 343\,K
for SCAN-rVV10, all consistent with prior molecular simulations
\cite{Potoff2001, Mathur2023, Goel2018}.  Compared to experiment, for both water (647\,K) and carbon dioxide (304\,K) \cite{NISTWebBook}, the empirical force fields give the best agreement. Among the xc functionals investigated, SCAN provides the
  best agreement for water, and PBE-D3 for carbon dioxide.

\textbf{\emph{Confined fluids}}. In the case of liquid--vapor
coexistence, inhomogeneity arises naturally at the interface between
phases. Away from coexistence, external potentials, such as those from
confining boundaries, can induce average inhomogeneous structure in
the fluid. To demonstrate the ability of \emph{ab initio} neural cDFT
to accurately describe confined fluids, we take $V_{\rm ext}$ as
two 9--3 Lennard--Jones potentials separated by $H$,
fitted to quantum Monte Carlo data for either a single water or carbon
dioxide molecule at a graphene surface
\cite{Brandenburg2019,DellaPia2025} (see \emph{Methods}).

As seen in Fig.~\ref{fig2}D, where we show $\rho(z)$ for confined
water described by the SCAN xc functional, \emph{ab initio} neural
cDFT is in excellent agreement with results from molecular dynamics
simulations. Note that, in these calculations, we solved the
Euler--Lagrange equation (Eq.~\ref{eqn:EL2}) subject to the constraint
that $\bar{\rho}_L$ matches the canonical simulations, and the
reported values for $\mu$ reflect the resulting Lagrange
multiplier.

Figure~\ref{fig2}D also demonstrates that thermodynamic
  properties under confinement are well-described. In
particular, for supercritical carbon dioxide at $T=400\,$K, we have
computed a measure of the effective pressure---the total force per
unit area $A$ exerted by the fluid on the confining walls---from the
derivative of the grand potential with respect to separation between
the graphene sheets,
\begin{equation}
  \begin{split}
    \label{eqn:Ptilde}
    \tilde{P} &\equiv -\frac{1}{A}\left(\frac{\partial\Omega}{\partial H}\right)_{A,T,\mu} = -\!\!\int^L_0\!\!\mrm{d}z\, \frac{\mrm{d}V_{\rm wall}(z)}{\mrm{d}z} \rho(z), 
  \end{split}
\end{equation}
where the right hand side represents a sum rule \cite{Evans1987}, with
$V_{\rm wall}$ the external potential of a single wall.  Following
directly from the standard thermodynamic treatment of inhomogeneous
fluids (see SI), the effective pressure $\tilde{P} = P + \Pi$
comprises the bulk pressure and the disjoining pressure $\Pi$, i.e.,
the additional mechanical contribution from the fluid required to
maintain the separation $H$ \cite{churaev2013surface}.  In order to
compare directly against GCMC simulations, results have been obtained
with the TraPPE empirical potential. Again, we observe good agreement
between the molecular simulations and the neural cDFT. Note that the
left and right hand sides of Eq.~\ref{eqn:Ptilde} provide two routes
to computing $\tilde{P}$; here we have presented results using the
latter, structural, option. Thermodynamic consistency between
the two approaches is assessed in the SI (Fig.~S8); it is
  overall very good above the critical point, with more pronounced
  discrepancies observed at subcritical temperatures.

\textbf{\emph{Training vs. validation}.}
We end this section with a brief comment on the extent to which
  the results presented so far should be regarded as interpolative
  reconstructions versus genuine out-of-sample predictions. The
  simulation data used to train \emph{ab initio} neural cDFT were
  obtained from canonical MD simulations across a range of
  temperatures, either in the presence of random $V_{\rm ext}(z)$ or
  at direct liquid--vapor coexistence. For the binodals shown in
  Fig.~\ref{fig2}C, neural cDFT has therefore been exposed to
  representative configurations from simulation. However, we note that
  cDFT minimization can be performed at any temperature across the
  temperature range, demonstrating its ability to interpolate smoothly
  between discrete training points (see Fig.~S5).

Importantly, no bulk simulation data---whether structural or
  thermodynamic (e.g., free energies, pressures, or chemical
  potentials)---were provided during training. The bulk equations of
  state and structure factors shown in Fig.~\ref{fig2}B should
  therefore be regarded as genuine held-out predictions, validated
  directly against the MLIP simulations. Similarly, the values of
  $\mu$ reported in Fig.~\ref{fig2}\tcr{D} are not known from MD simulations
  and are instead predictions of neural cDFT. While direct validation
  of these quantities against MLIP simulations is challenging, the
  excellent structural agreement between \emph{ab initio} neural cDFT
  and MD provides strong indirect support. The effective pressure
  under confinement, also shown in Fig.~\ref{fig2}D, is likewise not
  known \emph{a priori}. The predictive capability of \emph{ab initio}
  neural cDFT beyond its training set is perhaps most clearly
  illustrated by the fact that $c^{(1)}(z;[\rho],T)$---an object
  defined in the grand canonical ensemble---is learned entirely from
  data generated in canonical MD simulations.

\begin{figure*}[t]
\includegraphics[width=0.9\linewidth]{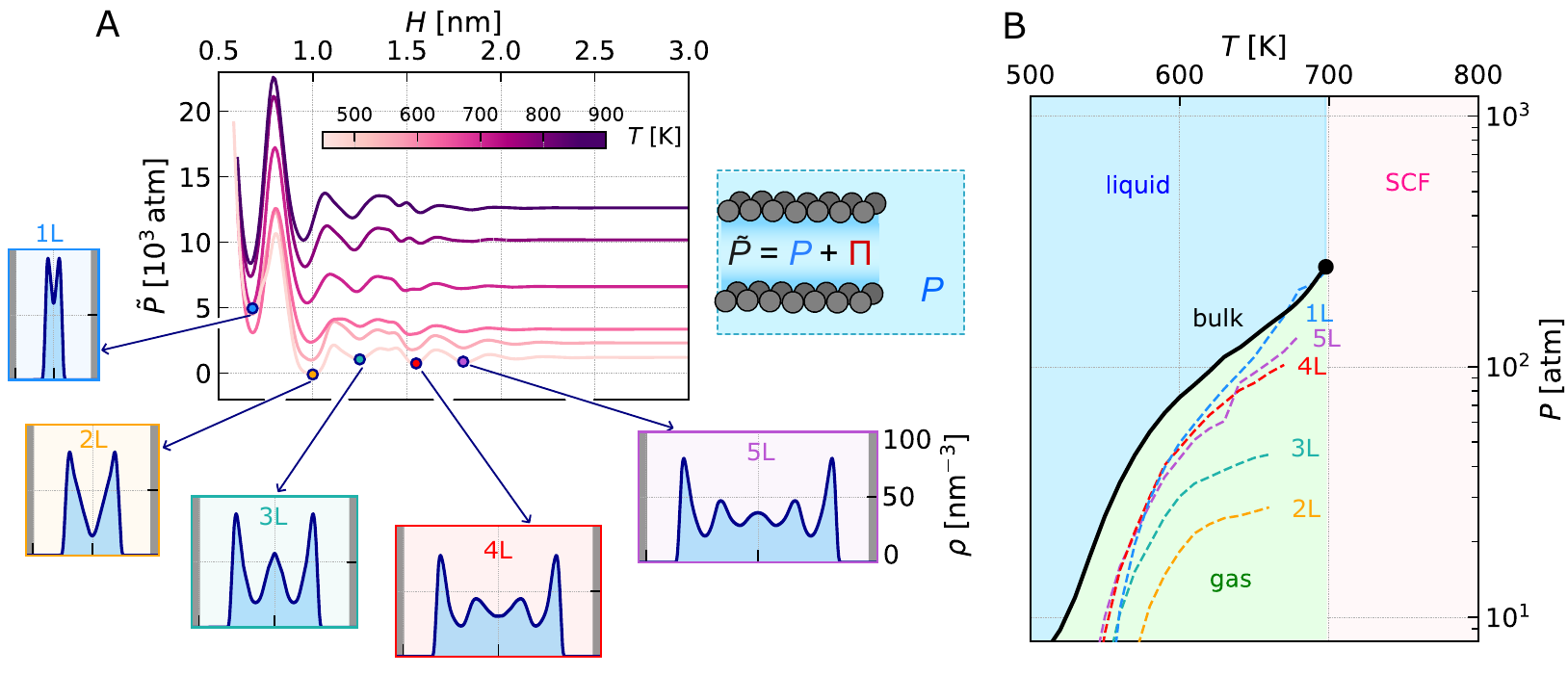}
\caption{\textbf{Predicting liquid--vapor equilibria of SCAN water
    upon confinement between graphene sheets.} A: The effective
  pressure $\tilde{P}$ (see Eq.~\ref{eqn:Ptilde}), for a graphene
  slit pore in equilibrium with a reservoir
  ($\rho_{\rm b} = 33$\,nm$^{-3}$) at different $T$. $\tilde{P}$
  exhibits minima at different $H$, each corresponding to a different
  number of water layers, as seen in the accompanying density profiles
  (the dark shaded regions indicate the positions of the graphene
  sheets). B: Liquid--vapor phase diagram in the $P$--$T$ plane, for
  the different $H$ indicated. Note that $P$ is the bulk pressure of
  the reservoir.}
\label{fig3}
\end{figure*}

\section*{Liquid--vapor coexistence of nanoconfined water}

Understanding how confinement influences the behavior of fluids has
garnered significant recent interest driven, in part, by advances in
fabricating nanofluidic channels \cite{Radha2016}. In the case of
water, recent experiments have reported a profound impact of extreme
confinement on its dielectric properties \cite{Fumagalli2018,
  Wang2025} and capillary behavior \cite{Yang2020, Zou2021}. Many
groups have also studied nanoconfined water with molecular
simulations, both with traditional empirical potentials and MLIPs
\cite{Han2010, Takaiwa2008, Algara-Siller2015, Kapil2022, Bui2023,
  Bui2024, Thiemann2022}. Importantly, not only are molecular
simulations used to cast light on experimental findings, they also act
as a predictive tool---a prime example is the prediction of new phases
under extreme confinement \cite{Kapil2022}.

When dealing with emergent phenomena such as phase behavior, a
consistent thermodynamic description is imperative to fully maximize
the predictive potential of molecular modeling \cite{Chew2023}.  For
typical molecular simulations, however, how to appropriately discuss
relevant thermodynamic variables under confinement can be complicated.
For example, a common approach for computing pressure under confinement
is to use the lateral components of the virial pressure tensor
\cite{Han2010, Takaiwa2008, Algara-Siller2015, Kapil2022}. For extreme
confinement, however, this measure of pressure depends sensitively on
the precise definition of the separation, which is, to a certain
  extent, arbitrary. In contrast, both $P$ (the pressure of the
reservoir) and $\Pi$ (the disjoining pressure) are insensitive to the
precise definition of $H$, and are readily obtainable within a
  cDFT formalism. We refer the reader to Ref.~\onlinecite{Evans1987},
and we also provide a brief overview in the SI.

Our results in Fig.~\ref{fig2}D already demonstrate the capability of
\emph{ab initio} neural cDFT to describe the structure and
thermodynamics of confined fluids. We now capitalize upon both its
conceptual and computational advantages to predict, from first
principles, how confinement influences water's liquid--vapor
coexistence. In Fig.~\ref{fig3}A, we present $\tilde{P}$ vs $H$ for
the graphene slit pore in a liquid state in equilibrium
  with a reservoir at $\rho_{\rm b}=33$\,nm$^{-3}$ at 
different $T$. As expected, in all cases we observe
$\lim_{H\to\infty}\tilde{P}\to P$. Strikingly, for $H\lesssim 2$\,nm,
we observe several local minima; these correspond to slit widths that
are commensurate with a particular number of layers of water, as can
be seen from visual inspection of the density profiles.

While we have shown results for a liquid-like state, under certain
conditions, the Euler--Lagrange equation can simultaneously admit a
solution corresponding to a vapor-like state; the stable phase is that
with lowest $\Omega$. To construct coexistence curves under
confinement, we therefore seek, for a given $H$ and $T$, the value of
$\mu$ at which the vapor- and liquid-like states have equal grand
potentials.  With molecular simulation, this
procedure---systematically varying $\mu$ to locate phase coexistence
across a range of $H$ and $T$---would require either many
GCMC simulations, which are impractical with MLIPs
\cite{Goeminne2023}, or an involved set of free energy
calculations. In contrast, with \emph{ab initio} neural
  cDFT the entire phase diagram is obtained in minutes. We present
the resulting confined liquid--vapor phase diagram in
Fig.~\ref{fig3}B. As the equation of state $P(\mu,T)$ is readily
obtainable from neural cDFT, to facilitate intuitive understanding, we
present results in the $P$--$T$ plane.

We immediately see from Fig.~\ref{fig3}B that, relative to bulk,
confinement acts to stabilize the liquid phase. In general, we also
observe that the critical temperature is shifted down as $H$
decreases, in line with expectations from the standard thermodynamic
treatment of inhomogeneous fluids \cite{Evans1987}. An exception to
this trend is the most extreme confinement corresponding to a single
layer of water, $H\approx 0.7$\,nm. As can be deduced from
Fig.~\ref{fig3}A, even though for a given bulk pressure the liquid is
still generally stabilized, under this extreme confinement $\Pi>0$;
this indicates that the fluid exerts a repulsive force on the
confining graphene sheets. This effect is made more apparent in
Fig.~S9, where we show the phase diagram in the $\tilde{P}$--$T$
plane.

\section*{Supercritical crossover lines in  carbon dioxide}

So far, we have shown \emph{ab initio} neural cDFT's ability to probe
liquid--vapor coexistence both in bulk, and under confinement. We now
demonstrate its potential to describe the physics of bulk fluids away
from coexistence, focusing on supercritical carbon dioxide
(sc\ce{CO2}). Understanding sc\ce{CO2}, and supercritical fluids
(SCFs) more generally, is not only important from a practical
viewpoint---it underpins technologies for carbon capture, sustainable
power generation, and chemical extraction \cite{White2021,
  Hsiao2025}---but also of fundamental interest. For example, fluids under
extreme conditions are often encountered deep in the interior of giant
``gas'' planets, and supercritical fluids are known to exhibit
nontrivial thermodynamic and dynamic behavior \cite{Pan2013, Pan2016}.

Here, we focus on the PBE-D3 xc functional, as we have already
established that it reasonably well describes carbon dioxide's
liquid--vapor coexistence, while results for other functionals
  are shown in Fig.~S7. In Fig.~\ref{fig4}A, we recast the phase
diagram in the $P$--$T$ plane, where we also compare directly to the
experimental result. While small differences between theory and
experiment are observed, agreement is overall very good. Turning our
attention to the supercritical state, we probe the bulk properties of
sc\ce{CO2} by analyzing the total correlation function $h(r)$, whose
Fourier transform is related to the two-body direct correlation
function of the uniform fluid, via the Ornstein--Zernike equation
\cite{HansenMcDonaldBook},
\begin{equation}
  \label{eqn:hk-OZ}
  \hat{h}(k) = \frac{\hat{c}^{(2)}_r(k)}{1-\rho_{\rm b}\hat{c}^{(2)}_r(k)}.
\end{equation}

For fluids with short-ranged interatomic potentials, it is well
established \cite{Evans1993} that, for a given temperature, the
asymptotic behavior of $h(r)$ changes from monotonic exponential at
low densities, i.e., \[h(r)\sim\exp(-\alpha_0 r),\] to oscillatory
with an exponential envelope, i.e.,
\[h(r)\sim\cos(\tilde{\alpha}_1 r-\theta)\exp(-\tilde{\alpha}_0 r).\]
The density at which this crossover occurs is the Fisher--Widom (FW)
transition \cite{Fisher1969, Evans1993}, and reflects the competition
between slowly-varying attractive interactions and rapidly-varying
repulsive forces that govern the packing of molecules; the locus
of points in the $\rho_{\rm b}$--$T$ plane at which this transition
occurs is the FW line.

\begin{figure*}[t!]
  \includegraphics[width=0.9\linewidth]{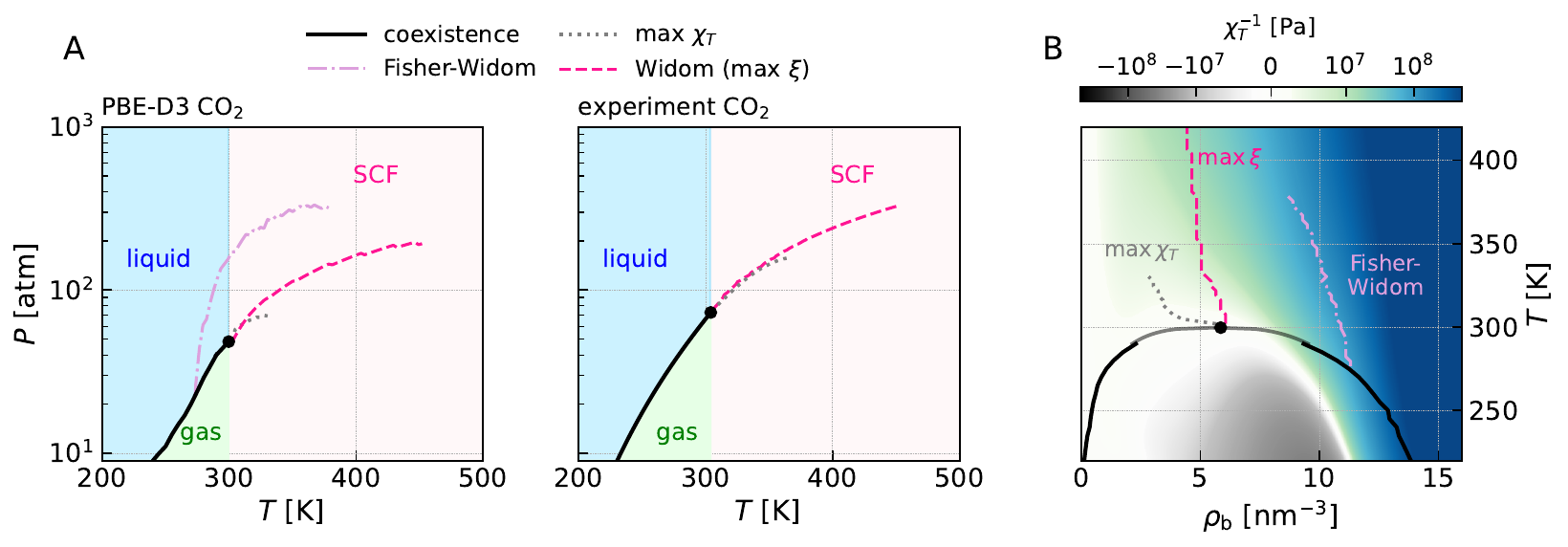}
  \caption{\textbf{Predicting behavior of supercritical PBE-D3 carbon
      dioxide.} A: $P$--$T$ phase diagram obtained with \emph{ab
      initio} neural cDFT (left) and experiment (right)
    \cite{Span1996, ProctorBook, nist_webbook}; overall, good
    agreement between the two is observed. B: $\rho_{\rm b}$--$T$ phase
    diagram with $\chi_{T}^{-1}$ superimposed as a heat map (see
    Eq.~\ref{eqn:chiT}). The Widom line obtained by $\max\chi_T$ is
    shown by the dotted line. We also show the Widom line obtained
    from $\max\xi$, where $\xi$ is the true correlation length (dashed
    line). The dot-dashed line shows the Fisher--Widom line,
    indicating a crossover from simple exponential to oscillatory
    asymptotic decay of the total correlation function. The Widom and
    Fisher--Widom lines are also plotted in panel A.}
\label{fig4}
\end{figure*}

As determining the FW line requires precise characterization of the
long-range behavior of $h(r)$, it is extremely challenging to obtain
from molecular simulation; accurate determination of the asymptotic
decay requires system sizes and sampling times that are
computationally prohibitive, particularly for \emph{ab initio}
descriptions. Consequently, previous studies have been limited to
simple model fluids \cite{Evans1993, Vega1995, Dijkstra2000}. With
\emph{ab initio} neural cDFT, the inverse decay lengths, $\alpha_0$
and $\tilde{\alpha}_0$, and the period $2\pi/\tilde{\alpha}_1$ can be
obtained directly from the pole structure of Eq.~\ref{eqn:hk-OZ} (also
see \emph{Methods}); the FW transition is then determined from the
density consistent with $\alpha_0 = \tilde{\alpha}_0$. In
Fig.~\ref{fig4}B, we plot carbon dioxide's FW line in the
$\rho_{\rm b}$--$T$ plane. In line with previous work for the
truncated-and-shifted LJ potential \cite{Evans1993}, we find that the
FW line intersects the binodal at $T/T_{\rm c}\approx 0.90$ and
$\rho_{\rm b}/\rho_{\rm c} \approx 1.93$.

A complementary description of the supercritical fluid is offered by
the ``Widom line'' \cite{Simeoni2010, Xu2005, Gallo2014,
  McMillan2010}, which was originally defined by the locus of points
in the $\rho_{\rm b}$--$T$ plane of maximum correlation length, though
it is typically associated with maxima in thermodynamic response
functions \cite{Fomin2015, Gorelli2006, Pipich2018, Li2024,
  Cockrell2021}. As shown in Ref.~\cite{Sammuller2025} for a LJ fluid,
the Widom line can be obtained from the maximum in the true
correlation length $\xi = 1/\alpha_0$, which we show in
Fig.~\ref{fig4}B. We also present the Widom line obtained from the
maximum in isothermal compressibility,
\begin{equation}
  \label{eqn:chiT}
  \chi_T(\rho_{\rm b},T) = \frac{\beta}{\rho_{\rm b}} S(k=0; \rho_{\rm b},T),
\end{equation}
which deviates from the Widom line obtained from $\max \xi$, in a
manner similar to that observed for the LJ fluid. Both sets of Widom
lines are also plotted on the $P$--$T$ phase diagrams in
Fig.~\ref{fig4}A. In this case, the discrepancy between the two
is less pronounced, with the Widom line computed from $\max \chi_{T}$
lying slightly below that obtained from $\max \xi$.

\section*{Discussion}

In this work, we have introduced \emph{ab initio} neural cDFT as a
simple, first-principles multiscale modeling framework for fluids. A
distinctive aspect of \emph{ab initio} neural cDFT compared to typical
multiscale methods is that it treats physics across length scales
within the same theoretical framework. In addition to obtaining the
equilibrium structure of the fluid, a key appealing feature of the
approach is its treatment of thermodynamics, for both homogeneous and
inhomogeneous systems.  In particular, as neural cDFT is formulated in
the grand canonical ensemble, the chemical potential of the fluid is
known, and measures of pressure in confined systems are well defined
and readily obtainable. We capitalized upon this feature to obtain the
liquid--vapor coexistence curves of water confined between graphene
sheets, with interatomic interactions described by the SCAN xc
functional. We also obtained the Fisher--Widom and Widom lines in
supercritical carbon dioxide, with interatomic interactions described
by the PBE-D3 xc functional; this demonstrates the framework's ability
to even describe bulk properties that would be challenging to obtain
from molecular simulations alone.

It is important to stress that our aim has been to introduce a
framework that provides accurate predictions of a fluid's emergent
physics equipped only with knowledge of its intermolecular
interactions, as determined by the Schr\"{o}dinger equation; we have
not made any attempt to improve the necessary approximations to the
many-body electronic structure problem. In this spirit, to the
  extent that the MLIPs provide suitable surrogate models, comparison
to experimental data should be viewed as informing on the
appropriateness of the underlying interatomic potential for the
problem at hand, rather than an assessment of the multiscale framework
itself. Whether \emph{ab initio} neural cDFT can be useful for
improving or validating the underlying electronic structure
approximation, e.g., by extending validation metrics to more
mesoscopic properties, remains an interesting question that lies
beyond the scope of the current study.

In the same vein, we have also used MLIPs that are readily available
\cite{Mathur2023, Wohlfahrt2020, Sanchez-Burgos2023}, rather than
trying to improve their description of the potential energy surface;
we have neither generated additional training data, nor developed a
new MLIP architecture. This means we have used MLIPs that rely on an
atom's local environment to determine the force acting upon it;
explicit electrostatic interactions are missing. Such neglect of
  long-ranged interactions can have important consequences for
  dielectric fluids such as water. In bulk, simulations reveal that
  the subsequent lack of screening causes long wavelength polarization
  fluctuations that differ from expected behavior, while for
  interfacial systems, erroneous polarization gradients can be
  observed \cite{yeh1999ewald, Rodgers2008pnas, Cox2020, Gao2022,
    Cheng2025}; a recent study of water at a copper substrate has even
  found that short-ranged MLIPs can lead to false metallization of the
  liquid layer \cite{parker2026false}.

Even for carbon dioxide, we cannot rule out the possibility that
the Fisher--Widom line reported in Fig.~\ref{fig4} is a
  consequence of the local description provided by the MLIP; the
  asymptotic behavior of correlations may be influenced by both
  interactions between permanent quadrupoles and London dispersion
  forces \cite{deCarvalho1994, evans2009pair}. Even if this is the
  case, the results discussed here likely provide a useful starting
  point for understanding the physics of the supercritical state. We
  also emphasize that our aim is to demonstrate how \emph{ab initio}
  neural cDFT can be used to assess the statistical mechanics of the
  underlying intermolecular potential in ways that can be challenging
  for molecular simulations. Clearly, a completely rigorous
description ought to take the effects of long-ranged electrostatics
explicitly into account, and we note recent progress from several
groups in this regard \cite{Yue2021, Niblett2021, Gao2022, Zhang2022,
  Ko2021, Ji2025, Cheng2025, Batatia2026, Kim2025, Loche2025,
  Caruso2026}.

Yet, by and large, for the properties that we have investigated,
it is likely that this neglect of long-ranged correlations is not too
severe. In particular, previous studies comparing long-ranged and
  short-ranged MLIPs have found that liquid--vapor density profiles
  are virtually indistinguishable \cite{Niblett2021, Cheng2025}. In a
  careful analysis, using ideas from local molecular field theory
  (LMFT) \cite{Rodgers2008, Rodgers2008pnas, Remsing2016, Cox2020},
  Niblett \emph{et al.} showed that short-ranged MLIPs attempt to
  learn the effects of long-ranged interactions implicitly, albeit in
  an uncontrolled manner \cite{Niblett2021}. With the assumption
that $c^{(1)}$ only depends on nearby values of the density, locality
also underpins the machine learning approach used in neural
cDFT\tcr{; it is this locality that enables predictions on length scales far 
beyond the reach of molecular simulations (see Figs.~1 and S2)} \cite{Sammuller2023}. 
Recently, building on ideas from LMFT, we have shown how the
effects of long-ranged electrostatics can be accurately accounted for
using a well-controlled mean-field approximation
\cite{Bui2025prl}. With the ability to model inhomogeneous
  systems over large length scales---where subtle effects of
  long-ranged correlations are likely to become pronounced---neural
  cDFT potentially provides a means to test MLIPs in ways that would
  be challenging with molecular simulations alone.

For molecular liquids, accounting for electrostatic interactions
  with neural cDFT relies on hyper-DFT, an extension of traditional
cDFT that permits the description of equilibrium observables
\cite{Sammuller2024hyper, bui2025hyperlmft}; in
  Refs.~\onlinecite{bui2025dielectro}
  and~\onlinecite{bui2025hyperlmft}, we treated the charge density as
  the observable. But hyper-DFT can be employed to access other
  observables that might be of interest. For example, in Fig.~S10, we
show how the density profile of confined water's hydrogen atoms can be
obtained, in addition to the one-body density prescribed by the oxygen
atoms. Importantly, the hyper-DFT framework ensures that
\emph{ab initio} neural cDFT can readily accommodate MLIPs that
explicitly treat long-ranged electrostatics \cite{bui2025hyperlmft},
and more generally, offers a route to formulating new theoretical
descriptions of fluid behavior. For example, we recently used
hyper-DFT to investigate how electric field gradients influence the
capillarity of dielectric fluids \cite{bui2025dielectro}.

As we acknowledged earlier, aside from the structure of the bulk
fluid, the neural cDFT we trained is limited to describing
  systems with planar inhomogeneous geometries. For the confined
  systems we have studied, this limitation means that the external
  potential approximates the effective interaction between the
  substrate and the fluid. Previous simulation studies that have
  directly compared such implicit substrate potentials against
  explicit atom representations suggest that many of the most salient
  aspects of confinement are well-described \cite{Kapil2022,
    chen2016two}. Even in cases where an explicit atom representation
  of the substrate is important, \emph{ab initio} neural cDFT may
  prove useful in providing a reasonable estimate of the system's
  thermodynamic state. \emph{Ab initio} neural cDFT is also limited
to describing the structure and thermodynamics of fluids at
equilibrium. For overdamped dynamics, it is possible to apply a
similar machine learning approach to non-equilibrium scenarios
\cite{Zimmermann2024}, but the applicability to molecular systems is
questionable. Even with these limitations in mind, the results we
present make clear that \emph{ab initio} neural cDFT is in a position
to cast light onto the influence of complex microscopic interactions
on emergent properties of fluids, and represents a significant
advancement in the \emph{ab initio} multiscale modeling of liquids.

Looking forward, one can envisage extending our strategy into a fully
machine-learned hierarchy spanning multiple physical scales. Recent
advances in machine-learned xc functionals \cite{Kirkpatrick2021,
  Pederson2022, luise2026} could provide electronic structure data
from which MLIPs are derived, which in turn generate the data used to
construct neural cDFT. Such a pipeline would constitute a unified,
fully machine-learned multiscale framework linking electronic
structure to the emergent physics of liquids. In the meantime, 
the integration of MLIPs and neural cDFT represents a distinctly novel 
approach to multiscale modeling of fluids, in which both the thermodynamics 
of open systems and mesoscale structure become readily accessible, 
while retaining a direct link to the underlying first-principles 
Hamiltonian.

\section*{Data Availability}

The code used to train the models and perform neural cDFT calculations in this study is available on Github (https://github.com/annatbui/mlip-neuraldft). Example simulation inputs, the MLIP models, and the training data are available on Zenodo (https://zenodo.org/records/21245362).

\section*{Supporting information}

Supporting information includes simulations, machine learning details and additional supporting results.

\section*{Acknowledgements}
Via membership of the United Kingdom Car Parrinello (UKCP) consortium
(Grant No. EP/F036884/1) and of the UK’s HEC Materials Chemistry
Consortium funded by EPSRC (EP/X035859), this work used the ARCHER2 UK
National Supercomputing Service.  A.T.B. acknowledges funding from the
Oppenheimer Fund and Peterhouse College, University of Cambridge.
S.J.C. is a Royal Society University Research Fellow (Grant
No. URF\textbackslash R1\textbackslash 211144) at Durham
University. We are grateful to Christoph Schran for useful
  discussions concerning molecular simulations of confined water.

\section*{Author contributions}
A.T.B.: Conceptualization (equal); Investigation (equal); Writing - original draft (equal); Writing - review \& editing (equal). S.J.C.: Conceptualization (equal); Investigation (equal); Writing - original draft (equal); Writing - review \& editing (equal)

{\small
\setlength{\parindent}{0pt}
\setlength{\parskip}{0.6em}
\section*{Methods}

\textbf{Machine-learned interatomic potentials.} For carbon dioxide,
the training datasets for energies and forces were taken from
Ref.~\cite{Mathur2023}. We considered three xc functionals: PBE
\cite{Perdew1996}, BLYP \cite{Becke1988} with D3 dispersion
corrections \cite{Grimme2010}, and SCAN-rVV10 \cite{Sun2015,
  Peng2016}. For each xc functional we trained an MLIP using the
DeepMD architecture with the \texttt{DeePMD-kit} package
\cite{Wang2018}, minimizing the loss function
\begin{equation*}
  \begin{split}
    \mcl{L} & = p_E \sum^n_{k=1}\left| E_{\mrm{ref}}(\mbf{R}^N_k) - E(\mbf{R}^N_k)  \right|^2 \\
            & + p_f \sum^n_{k=1}\sum^N_{i=1}\left|\left| \mbf{f}_{i,\mrm{ref}}(\mbf{R}^N_k) - \mbf{f}_i(\mbf{R}^N_k) \right|\right|^2,
  \end{split}
\end{equation*}
where $p_E$ and $p_f$ are tunable parameters that varied during the
optimization, $\mbf{R}^N_k$ is the $k^{\rm th}$ atomic configuration
in the training set with atomic forces
$\mbf{f}_{i,\mrm{ref}}(\mbf{R}^N_k)$ and energy
$E_{\mrm{ref}}(\mbf{R}^N_k)$, while $\mbf{f}_i(\mbf{R}^N_k)$ and
energy $E(\mbf{R}^N_k)$ are the corresponding model predictions
($i = 1\ldots N$ indexes the atoms). For reasons of computational
efficiency, we employed DeepMD for the carbon dioxide MLIPs used in
the main article. However, we also trained an MLIP for PBE-D3 carbon
  dioxide using MACE to show the robustness of the method to different
  MLIP architectures (see Fig.~S1). For water, we considered two xc
functionals: RPBE \cite{Hammer1999} with D3 dispersion corrections,
and SCAN. For RPBE-D3 water, we used the MLIP trained in
Ref.~\cite{Wohlfahrt2020} based upon the HD-NNP architecture. For SCAN
water, we used the MLIP trained in Ref.~\cite{Sanchez-Burgos2023}
based upon the DeepMD architecture.

\textbf{Generation of training data for neural cDFT.}  To obtain
training data for the neural network representations of
$c^{(1)}(z; [\rho], T)$, for each fluid and intermolecular potential
we performed MD simulations under random $V_{\rm ext}$ of the form \cite{Sammuller2023}
\[
  V_{\mrm{ext}}(z) = \sum^4_{n=1} A_n \sin\left(\frac{2\pi n z}{L_z} + \theta_n\right) + \sum^4_{n=1} B^{\mrm{lin}}_n(z)
\]
where $L_z$ is the simulation box length in the $z$ direction.  The
phases $\theta_n$ were chosen uniformly in the interval [0, $2\pi$),
and values of $A_n$ were drawn from an unbiased normal distribution
with variance of $0.4\,(k_{\mrm{B}}T)^2$.  The second summation
denotes up to four piecewise linear functions
\[
  B^{\mrm{lin}}_n(z) =  \begin{cases}
    V_1 + \dfrac{V_2 - V_1}{z_2 - z_1}(z - z_1) & \quad   z_1 < z < z_2\\
    0  & \quad  \text{otherwise},
  \end{cases}  
\]
with $0 < z_1 < z_2 < L_z$.  The locations $z_1$ and $z_2$ were
distributed uniformly while $V_1$ and $V_2$ were chosen randomly from
an unbiased normal distribution with variance of
$0.8\,(k_{\mrm{B}}T)^2$. From these randomized $V_{\mrm{ext}}$, the
external force $f_{\rm ext}(z) = -\partial_z V_{\rm ext}(z)$ was
obtained by finite difference and added to the molecular centers
(i.e., the carbon atom for carbon dioxide, and the oxygen atom for
water). In some cases, planar walls
of the form of a 9-3 Lennard--Jones potential,
\[
  V_{\rm wall}(z) = \epsilon_{\rm wall}
  \left[\frac{2}{15}\bigg(\frac{\sigma_{\rm wall}}{z}\bigg)^9
    - \bigg(\frac{\sigma_{\rm wall}}{z}\bigg)^3\right]
\]
were also included, with $\epsilon_{\rm wall}\in$ [$0.02$,
$2$]$\,k_{\mrm{B}}T$ and $\sigma_{\rm wall}\in[0.1,0.3]\,$nm.

Simulation cells of size $3.332\times3.332\times3.332\,\mrm{nm}^3$
for carbon dioxide and $2.000\times 2.000\times 4.000\,\mrm{nm}^3$ for
water were used, and periodic boundary conditions were employed. MD
simulations were performed in the canonical ensemble, from which the
density profiles were sampled after at least 1\,ns of production
run. The number of molecules in the box was randomized between 70--512
for carbon dioxide and 40--1024 for water. For TraPPE carbon dioxide,
the linear molecules are evolved as rigid bodies
\cite{Miller2002}. For TIP4P/2005 water, the geometries of the water
molecules were constrained using the \texttt{RATTLE} algorithm
\cite{Andersen1983}. Dynamics were propagated using the velocity
Verlet algorithm with a time-step of 0.5\,fs. The temperature, chosen
randomly between $220-500\,$K for carbon dioxide and $450-950\,$K for
water, was maintained using a Nos\'e--Hoover thermostat
\cite{Shinoda2004,Tuckerman2006}.  We used the \texttt{LAMMPS}
simulation package \cite{Thompson2022} interfaced with the
\texttt{n2p2} package \cite{Singraber2019} for the HD-NNP potential
and the \texttt{DeepMD-kit} package for the DeepMD
potentials. Overall, for each fluid and interatomic potential, we
sampled $500-1000$ simulations to generate the training data for neural
cDFT.

We also performed direct coexistence MD simulations with
  simulation cell sizes $3.332\times3.332\times20.000\,\mrm{nm}^3$ for
  carbon dioxide (comprising 1536 molecules) and
  $2.000\times2.000\times20.000\,\mrm{nm}^3$ for water (comprising 832
  molecules). For  each interatomic potential, approximately $20$ simulations were performed at random $T$ 
  from 220\,K up to $T_{\mrm{c}}$ for carbon dioxide, and from 450\,K up to $T_{\mrm{c}}$ for water.
  Density profiles were obtained after at least 1\,ns
  production run.

\textbf{Training neural cDFT}. To learn $c^{(1)}([\rho], T)$
for each fluid and interatomic potential, we employed a local learning
strategy detailed elsewhere \cite{Sammuller2023,sammuller2026}. In
brief, inputs consisted of local density in a sliding spatial window
of size $1\,\mrm{nm}$ from the center of the position of interest,
along with a separate input node that encodes $T$ as a scalar. The
loss function is defined based on Eq.~\ref{eqn:EL2} as
\[
  \mcl{L} = \sum^n_{k=1}\left|\left| \ln\big(\zeta^{-1}\Lambda^3\rho_k(z)\big) + \beta_{k} V^{(k)}_{\mrm{ext}}(z) -  \beta_{k}\mu_k - c^{(1)}(z;[\rho_k],T_k) \right|\right|^2
\]
where $k$ indexes each simulation in the dataset. Note that, as we
have used canonical MD simulations to generate our training data,
$\{\mu_k\}$ are treated as latent variables that are learned during
the training procedure. In practice, we set $\zeta^{-1}\Lambda^3 = 1\,\text{\AA}^{3}$.

The machine learning routine was implemented in Keras/Tensorflow
\cite{Chollet2017}. One fifth of the dataset was used for validation,
and the rest was used for training.  Models were trained for 100
epochs with a batch size of 128, using an exponentially decaying
learning rate starting at 0.001, achieving errors comparable to the
estimated simulation noise.  The network contains three
fully-connected layers, each containing 128, 64 and 32 nodes
respectively, with softplus activation. The training of the neural
networks was done on a GPU (\texttt{NVIDIA GeForce RTX 3060}) in a few
hours.

\textbf{Evaluating neural cDFT.} Evaluating the trained neural
functionals is fast ($\sim$ milliseconds) and can be performed on a
CPU or GPU.  The EL equation was solved self-consistently with a mixed
Picard iteration scheme, which typically converges within
minutes. When constraining $\bar{\rho}_L$, $\mu$ acts as the Lagrange
multiplier, which is achieved by a primal--dual gradient scheme.

To evaluate the excess free energy, we use functional line
  integration,
  \[
    \beta\mcl{F}^{\mrm{(ex)}}_{\mrm{intr}}([\rho],T)/\mcl{A} = -\int^1_0\!\!\mrm{d}\lambda\! \int\!\!\mrm{d} z\, \rho(z) \, c^{(1)}(z; [\lambda\rho], T),
  \]
where $\mcl{A}$ is the cross-sectional area.

To obtain the Widom and Fisher--Widom lines, a pole analysis was
performed following Refs.~\cite{Evans1993,Sammuller2025}, i.e.,
by finding the zeros of the denominator $1-\rho \hat{c}_{\rm r}^{(2)}(\alpha)$
in Eq.~\ref{eqn:OZ}. Here, $\alpha(\rho_{\rm b},T)=\alpha_1+i\alpha_0$,
  with $\alpha_0$ and $\alpha_1$ satisfying
\[
1=4\pi\rho_{\rm b}\!\int^\infty_0\!\!\mrm{d}r\,r^2 c_{\rm r}^{(2)}(r;[\rho_{\rm b}],T)\frac{\sinh(\alpha_0 r )}{\alpha_0 r } \cos(\alpha_1 r),
\]
\[
1=4\pi\rho_{\rm b}\!\int^\infty_0\!\!\mrm{d}r\,r^2 c_{\rm r}^{(2)}(r;[\rho_{\rm b}],T) \cosh(\alpha_0 r) \frac{\sin(\alpha_1 r )}{\alpha_1 r }.
\]

\textbf{Validation simulations.}  For the simulation data in
Fig.~\ref{fig2},  to
obtain the pressure and structure factors, we performed simulations of
the bulk fluid with simulation cells
$3.332\times3.332\times3.332\,\mrm{nm}^3$ for carbon dioxide, and
$2.000\times2.000\times2.000\,\mrm{nm}^3$ for water.
Direct coexistence simulations were performed as described above (see \textbf{Generation of training data for neural cDFT}).
For simulations
of the confined fluids, we employed Lennard--Jones 9-3 walls (with a
$1$\,nm cutoff), separated by $H$. To avoid interactions between
periodic images, the box dimension $L_z$ was chosen such that
$L_z - H = 4\,\mrm{nm}$. The wall--fluid interaction parameters were
obtained by fitting to quantum Monte Carlo data \cite{Brandenburg2019,
  DellaPia2025}, yielding $\epsilon_{\rm wall}=0.1298\,\mrm{eV}$,
$\sigma_{\rm wall}=0.3868\,\mrm{nm}$ for carbon dioxide and
$\epsilon_{\rm wall}=0.0829\,\mrm{eV}$,
$\sigma_{\rm wall}=0.3590\,\mrm{nm}$ for water. When benchmarking the
disjoining pressure of TraPPE carbon dioxide, we performed GCMC
simulations with our own code \cite{githubgcmc}.  }

\bibliography{references}

@article{Shinoda2004,
  title = {Rapid estimation of elastic constants by molecular dynamics simulation under constant stress},
  author = {Shinoda, Wataru and Shiga, Motoyuki and Mikami, Masuhiro},
  journal = {Phys. Rev. B},
  volume = {69},
  issue = {13},
  pages = {134103},
  numpages = {8},
  year = {2004},
  month = {Apr},
  publisher = {American Physical Society},
  doi = {10.1103/PhysRevB.69.134103},
  url = {https://link.aps.org/doi/10.1103/PhysRevB.69.134103}
}

@Article{Reuter2006,
author={Reuter, Karsten
and Scheffler, Matthias},
  title = {First-principles kinetic {Monte Carlo} simulations for heterogeneous catalysis: Application to the {CO} oxidation at $\mathrm{Ru}$$\mathrm{{O}}_{2}(110)$},
journal={Phys. Rev. B},
year={2006},
month={Jan},
day={26},
publisher={American Physical Society},
volume={73},
number={4},
pages={045433},
doi={10.1103/PhysRevB.73.045433},
url={https://doi.org/10.1103/PhysRevB.73.045433}
}

@Article{Schran2021,
author={Schran, Christoph
and Thiemann, Fabian L.
and Rowe, Patrick
and M{\"u}ller, Erich A.
and Marsalek, Ondrej
and Michaelides, Angelos},
title={Machine learning potentials for complex aqueous systems made simple},
journal={Proc. Natl. Acad. Sci. U.S.A},
year={2021},
month={Sep},
day={21},
publisher={Proc. Natl. Acad. Sci. U.S.A},
volume={118},
number={38},
pages={e2110077118},
doi={10.1073/pnas.2110077118},
url={https://doi.org/10.1073/pnas.2110077118}
}

@Article{DellaPia2025,
author={Della Pia, Flaviano
and Kler-Young, Giaan
and Zen, Andrea
and Berger, Fabian
and Alf{\`e}, Dario
and Michaelides, Angelos},
title={Interaction strength of carbon dioxide on graphene from periodic quantum diffusion {Monte Carlo}},
journal={J. Chem. Phys},
year={2025},
month={Aug},
day={19},
volume={163},
number={7},
pages={071101},
issn={0021-9606},
doi={10.1063/5.0283254},
url={https://doi.org/10.1063/5.0283254}
}

@Article{Brandenburg2019,
author={Brandenburg, Jan Gerit
and Zen, Andrea
and Fitzner, Martin
and Ramberger, Benjamin
and Kresse, Georg
and Tsatsoulis, Theodoros
and Gr{\"u}neis, Andreas
and Michaelides, Angelos
and Alf{\`e}, Dario},
title={Physisorption of Water on Graphene: Subchemical Accuracy from Many-Body Electronic Structure Methods},
journal={J. Phys. Chem. Lett},
year={2019},
month={Feb},
day={07},
publisher={American Chemical Society},
volume={10},
number={3},
pages={358-368},
doi={10.1021/acs.jpclett.8b03679},
url={https://doi.org/10.1021/acs.jpclett.8b03679}
}

@Article{Andersen1983,
author={Andersen, Hans C.},
title={Rattle: A ``velocity'' version of the shake algorithm for molecular dynamics calculations},
journal={J. Comput. Phys},
year={1983},
month={Oct},
day={01},
volume={52},
number={1},
pages={24-34},
issn={0021-9991},
doi={10.1016/0021-9991(83)90014-1}
}

@misc{NISTWebBook,
  author       = {P. J. Linstrom and W. G. Mallard},
  title        = {{NIST} Chemistry WebBook, {NIST} Standard Reference Database Number 69},
  howpublished = {National Institute of Standards and Technology, Gaithersburg, MD 20899},
  year         = {2026},
  doi          = {10.18434/T4D303},
  note         = {Retrieved March 20, 2026}
}

@Article{Cheng2019,
author={Cheng, Bingqing
and Engel, Edgar A.
and Behler, J{\"o}rg
and Dellago, Christoph
and Ceriotti, Michele},
title={Ab initio thermodynamics of liquid and solid water},
journal={Proc. Natl. Acad. Sci. U.S.A},
year={2019},
month={Jan},
day={22},
publisher={Proc. Natl. Acad. Sci. U.S.A},
volume={116},
number={4},
pages={1110-1115},
doi={10.1073/pnas.1815117116},
url={https://doi.org/10.1073/pnas.1815117116}
}

@Article{Karplus2002,
author={Karplus, Martin
and McCammon, J. Andrew},
title={Molecular dynamics simulations of biomolecules},
journal={Nat. Struct. Mol. Biol},
year={2002},
month={Sep},
day={01},
volume={9},
number={9},
pages={646-652},
issn={1545-9985},
doi={10.1038/nsb0902-646},
url={https://doi.org/10.1038/nsb0902-646}
}

@Article{Field1990,
author={Field, Martin J.
and Bash, Paul A.
and Karplus, Martin},
title={A combined quantum mechanical and molecular mechanical potential for molecular dynamics simulations},
journal={J. Comput. Chem},
year={1990},
month={Jul},
day={01},
publisher={John Wiley {\&} Sons, Ltd},
volume={11},
number={6},
pages={700-733},
issn={0192-8651},
doi={10.1002/jcc.540110605},
url={https://doi.org/10.1002/jcc.540110605}
}

@Article{Bruix2019,
author={Bruix, Albert
and Margraf, Johannes T.
and Andersen, Mie
and Reuter, Karsten},
title={First-principles-based multiscale modelling of heterogeneous catalysis},
journal={Nat. Catal},
year={2019},
month={Aug},
day={01},
volume={2},
number={8},
pages={659-670},
issn={2520-1158},
doi={10.1038/s41929-019-0298-3},
url={https://doi.org/10.1038/s41929-019-0298-3}
}

@article{Tuckerman2006,
	doi = {10.1088/0305-4470/39/19/s18},
	url = {https://doi.org/10.1088/0305-4470/39/19/s18},
	year = 2006,
	month = {apr},
	publisher = {{IOP} Publishing},
	volume = {39},
	number = {19},
	pages = {5629--5651},
	author = {Mark E Tuckerman and Jos{\'{e}} Alejandre and Roberto L{\'{o}}pez-Rend{\'{o}}n and Andrea L Jochim and Glenn J Martyna},
	title = {A {Liouville-operator} derived measure-preserving integrator for molecular dynamics simulations in the isothermal{\textendash}isobaric ensemble},
	journal = {J. Phys. A: Math. Gen.}
}

@article{Thompson2022,
title = {{LAMMPS} - a flexible simulation tool for particle-based materials modeling at the atomic, meso, and continuum scales},
journal = {Comput. Phys. Commun},
volume = {271},
pages = {108171},
year = {2022},
issn = {0010-4655},
doi = {https://doi.org/10.1016/j.cpc.2021.108171},
url = {https://www.sciencedirect.com/science/article/pii/S0010465521002836},
author = {Aidan P. Thompson and H. Metin Aktulga and Richard Berger and Dan S. Bolintineanu and W. Michael Brown and Paul S. Crozier and Pieter J. {in 't Veld} and Axel Kohlmeyer and Stan G. Moore and Trung Dac Nguyen and Ray Shan and Mark J. Stevens and Julien Tranchida and Christian Trott and Steven J. Plimpton}
}

@book{Chollet2017,
  title={Deep {Learning} with {Python}},
  author={Chollet, F.},
  isbn={9781617294433},
  url={https://www.manning.com/books/deep-learning-with-python},
  year={2017},
  publisher={Manning Publications}
}

@Article{Sammuller2023,
author={Samm{\"u}ller, Florian
and Hermann, Sophie
and de las Heras, Daniel
and Schmidt, Matthias},
title={Neural functional theory for inhomogeneous fluids: {F}undamentals and applications},
journal={Proc. Natl. Acad. Sci. U.S.A},
year={2023},
month={Dec},
day={12},
publisher={Proc. Natl. Acad. Sci. U.S.A},
volume={120},
number={50},
pages={e2312484120},
doi={10.1073/pnas.2312484120},
url={https://doi.org/10.1073/pnas.2312484120}
}

@misc{githubgcmc,
  author = {Anna T. Bui},
  title = {{GCMC with Gaussian truncated potentials}},
  howpublished = "\url{https://github.com/annatbui/GCMC}",
  year = {2024}, 
}

@book{HansenMcDonaldBook,
  title={Theory of Simple Liquids: with Applications to Soft Matter},
  author={Hansen, J.P. and McDonald, I.R.},
  isbn={9780123870339},
  lccn={2013372077},
  year={2013},
  publisher={Elsevier Science}
}

@Article{Robitschko2025,
author={Robitschko, Silas
and Samm{\"u}ller, Florian
and Schmidt, Matthias
and Evans, Robert},
title={Learning the bulk and interfacial physics of liquid--liquid phase separation with neural density functionals},
journal={J. Chem. Phys},
year={2025},
month={Oct},
day={22},
volume={163},
number={16},
pages={161101},
issn={0021-9606},
doi={10.1063/5.0290261},
url={https://doi.org/10.1063/5.0290261}
}

@article{Bui2025prl,
  title = {Learning Classical Density Functionals for Ionic Fluids},
  author = {Bui, Anna T. and Cox, Stephen J.},
  journal = {Phys. Rev. Lett.},
  volume = {134},
  issue = {14},
  pages = {148001},
  numpages = {8},
  year = {2025},
  month = {Apr},
  publisher = {American Physical Society},
  doi = {10.1103/PhysRevLett.134.148001},
  url = {https://link.aps.org/doi/10.1103/PhysRevLett.134.148001}
}

@article{Sammuller2025,
  title = {Neural Density Functional Theory of Liquid-Gas Phase Coexistence},
  author = {Samm\"uller, Florian and Schmidt, Matthias and Evans, Robert},
  journal = {Phys. Rev. X},
  volume = {15},
  issue = {1},
  pages = {011013},
  numpages = {23},
  year = {2025},
  month = {Jan},
  publisher = {American Physical Society},
  doi = {10.1103/PhysRevX.15.011013},
  url = {https://link.aps.org/doi/10.1103/PhysRevX.15.011013}
}

@article{Sammuller2024hyper,
  title = {Hyperdensity Functional Theory of Soft Matter},
  author = {Samm\"uller, Florian and Robitschko, Silas and Hermann, Sophie and Schmidt, Matthias},
  journal = {Phys. Rev. Lett.},
  volume = {133},
  issue = {9},
  pages = {098201},
  numpages = {7},
  year = {2024},
  month = {Aug},
  publisher = {American Physical Society},
  doi = {10.1103/PhysRevLett.133.098201},
  url = {https://link.aps.org/doi/10.1103/PhysRevLett.133.098201}
}

@Article{bui2025hyperlmft,
author={Bui, Anna T.
and Cox, Stephen J.},
title={A first-principles approach to electromechanics in liquids},
journal={J. Phys.: Condens. Matter},
year={2025},
month={Jul},
day={08},
publisher={IOP Publishing},
volume={37},
number={28},
pages={285101},
issn={0953-8984},
doi={10.1088/1361-648X/ade7e7},
url={https://doi.org/10.1088/1361-648X/ade7e7}
}

@article{Sun2015,
  title = {Strongly Constrained and Appropriately Normed Semilocal Density Functional},
  author = {Sun, Jianwei and Ruzsinszky, Adrienn and Perdew, John P.},
  journal = {Phys. Rev. Lett.},
  volume = {115},
  issue = {3},
  pages = {036402},
  numpages = {6},
  year = {2015},
  month = {Jul},
  publisher = {American Physical Society},
  doi = {10.1103/PhysRevLett.115.036402},
  url = {https://link.aps.org/doi/10.1103/PhysRevLett.115.036402}
}

@article{Hammer1999,
  title = {Improved adsorption energetics within density-functional theory using revised {Perdew-Burke-Ernzerhof} functionals},
  author = {Hammer, B. and Hansen, L. B. and N\o{}rskov, J. K.},
  journal = {Phys. Rev. B},
  volume = {59},
  issue = {11},
  pages = {7413--7421},
  numpages = {0},
  year = {1999},
  month = {Mar},
  publisher = {American Physical Society},
  doi = {10.1103/PhysRevB.59.7413},
  url = {https://link.aps.org/doi/10.1103/PhysRevB.59.7413}
}

@Article{Singraber2019,
author={Singraber, Andreas
and Behler, J{\"o}rg
and Dellago, Christoph},
title={Library-Based {LAMMPS} Implementation of High-Dimensional Neural Network Potentials},
journal={J. Chem. Theory Comput},
year={2019},
month={Mar},
day={12},
publisher={American Chemical Society},
volume={15},
number={3},
pages={1827-1840},
issn={1549-9618},
doi={10.1021/acs.jctc.8b00770},
url={https://doi.org/10.1021/acs.jctc.8b00770}
}

@article{Behler2007,
  title = {Generalized Neural-Network Representation of High-Dimensional Potential-Energy Surfaces},
  author = {Behler, J\"org and Parrinello, Michele},
  journal = {Phys. Rev. Lett.},
  volume = {98},
  issue = {14},
  pages = {146401},
  numpages = {4},
  year = {2007},
  month = {Apr},
  publisher = {American Physical Society},
  doi = {10.1103/PhysRevLett.98.146401},
  url = {https://link.aps.org/doi/10.1103/PhysRevLett.98.146401}
}

@article{Zhang2018,
  title = {Deep Potential Molecular Dynamics: A Scalable Model with the Accuracy of Quantum Mechanics},
  author = {Zhang, Linfeng and Han, Jiequn and Wang, Han and Car, Roberto and E, Weinan},
  journal = {Phys. Rev. Lett.},
  volume = {120},
  issue = {14},
  pages = {143001},
  numpages = {6},
  year = {2018},
  month = {Apr},
  publisher = {American Physical Society},
  doi = {10.1103/PhysRevLett.120.143001},
  url = {https://link.aps.org/doi/10.1103/PhysRevLett.120.143001}
}

@inproceedings{Batatia2022,
 author = {Batatia, Ilyes and Kovacs, David P and Simm, Gregor and Ortner, Christoph and Csanyi, Gabor},
 booktitle = {Adv. Neural Inf. Process. Syst.},
 editor = {S. Koyejo and S. Mohamed and A. Agarwal and D. Belgrave and K. Cho and A. Oh},
 pages = {11423--11436},
 publisher = {Curran Associates, Inc.},
 title = {{MACE}: Higher Order Equivariant Message Passing Neural Networks for Fast and Accurate Force Fields},
 url = {https://proceedings.neurips.cc/paper_files/paper/2022/file/4a36c3c51af11ed9f34615b81edb5bbc-Paper-Conference.pdf},
 volume = {35},
 year = {2022}
}

@Article{Goel2018,
author={Goel, Himanshu
and Windom, Zachary W.
and Jackson, Amber A.
and Rai, Neeraj},
title={Performance of density functionals for modeling vapor liquid equilibria of {CO$_2$} and {SO$_2$}},
journal={J. Comput. Chem},
year={2018},
month={Mar},
day={30},
publisher={John Wiley {\&} Sons, Ltd},
volume={39},
number={8},
pages={397-406},
issn={0192-8651},
doi={10.1002/jcc.25123},
url={https://doi.org/10.1002/jcc.25123}
}

@Article{Goeminne2023,
author={Goeminne, Ruben
and Vanduyfhuys, Louis
and Van Speybroeck, Veronique
and Verstraelen, Toon},
title={{DFT}-Quality Adsorption Simulations in Metal--Organic Frameworks Enabled by Machine Learning Potentials},
journal={J. Chem. Theory Comput},
year={2023},
month={Sep},
day={26},
publisher={American Chemical Society},
volume={19},
number={18},
pages={6313-6325},
issn={1549-9618},
doi={10.1021/acs.jctc.3c00495},
url={https://doi.org/10.1021/acs.jctc.3c00495}
}

@Article{Rodgers2008pnas,
author={Rodgers, Jocelyn M.
and Weeks, John D.},
title={Interplay of local hydrogen-bonding and long-ranged dipolar forces in simulations of confined water},
journal={Proc. Natl. Acad. Sci. U.S.A},
year={2008},
month={Dec},
day={09},
publisher={Proc. Natl. Acad. Sci. U.S.A},
volume={105},
number={49},
pages={19136-19141},
doi={10.1073/pnas.0807623105},
url={https://doi.org/10.1073/pnas.0807623105}
}

@article{sammuller2026,
  title = {Determining the Chemical Potential via Universal Density Functional Learning},
  author = {Samm\"uller, Florian and Schmidt, Matthias},
  journal = {Phys. Rev. Lett.},
  volume = {136},
  issue = {6},
  pages = {068202},
  numpages = {9},
  year = {2026},
  month = {Feb},
  publisher = {American Physical Society},
  doi = {10.1103/7bqn-y2d7},
  url = {https://link.aps.org/doi/10.1103/7bqn-y2d7}
}

@Article{Zhang2024,
author={Zhang, Hanwen
and Juraskova, Veronika
and Duarte, Fernanda},
title={Modelling chemical processes in explicit solvents with machine learning potentials},
journal={Nat. Commun},
year={2024},
month={Jul},
day={20},
volume={15},
number={1},
pages={6114},
issn={2041-1723},
doi={10.1038/s41467-024-50418-6},
url={https://doi.org/10.1038/s41467-024-50418-6}
}

@book{rowlinson2002book,
  title={Molecular Theory of Capillarity},
  author={Rowlinson, J.S. and Widom, B.},
  isbn={9780486425443},
  lccn={2002074022},
  series={Dover books on chemistry},
  year={2002},
  publisher={Dover Publications}
}

@book{churaev2013surface,
  title={Surface Forces},
  author={Churaev, N.V. and Derjaguin, B.V. and Muller, V.M.},
  isbn={9781475766394},
  series={Chemistry and Materials Science},
  url={https://books.google.co.uk/books?id=ZdLkBwAAQBAJ},
  year={2013},
  publisher={Springer US}
}

@Article{Evans1993,
author={Evans, R.
and Henderson, J. R.
and Hoyle, D. C.
and Parry, A. O.
and Sabeur, Z. A.},
title={Asymptotic decay of liquid structure: oscillatory liquid-vapour density profiles and the {Fisher-Widom} line},
journal={Mol. Phys.},
year={1993},
month={Nov},
day={01},
publisher={Taylor {\&} Francis},
volume={80},
number={4},
pages={755-775},
issn={0026-8976},
doi={10.1080/00268979300102621},
url={https://doi.org/10.1080/00268979300102621}
}

@Article{Evans1987,
author={Evans, R.
and Marini Bettolo Marconi, U.},
title={Phase equilibria and solvation forces for fluids confined between parallel walls},
journal={J. Chem. Phys},
year={1987},
month={Jun},
day={15},
volume={86},
number={12},
pages={7138-7148},
issn={0021-9606},
doi={10.1063/1.452363},
url={https://doi.org/10.1063/1.452363}
}

@Article{Bui2023,
author={Bui, Anna T.
and Thiemann, Fabian L.
and Michaelides, Angelos
and Cox, Stephen J.},
title={Classical Quantum Friction at Water--Carbon Interfaces},
journal={Nano Lett},
year={2023},
month={Jan},
day={25},
publisher={American Chemical Society},
volume={23},
number={2},
pages={580-587},
issn={1530-6984},
doi={10.1021/acs.nanolett.2c04187},
url={https://doi.org/10.1021/acs.nanolett.2c04187}
}

@Article{Algara-Siller2015,
author={Algara-Siller, G.
and Lehtinen, O.
and Wang, F. C.
and Nair, R. R.
and Kaiser, U.
and Wu, H. A.
and Geim, A. K.
and Grigorieva, I. V.},
title={Square ice in graphene nanocapillaries},
journal={Nature},
year={2015},
month={Mar},
day={01},
volume={519},
number={7544},
pages={443-445},
issn={1476-4687},
doi={10.1038/nature14295},
url={https://doi.org/10.1038/nature14295}
}

@Article{Thiemann2022,
author={Thiemann, Fabian L.
and Schran, Christoph
and Rowe, Patrick
and M{\"u}ller, Erich A.
and Michaelides, Angelos},
title={Water Flow in Single-Wall Nanotubes: Oxygen Makes It Slip, Hydrogen Makes It Stick},
journal={ACS Nano},
year={2022},
month={Jul},
day={26},
publisher={American Chemical Society},
volume={16},
number={7},
pages={10775-10782},
issn={1936-0851},
doi={10.1021/acsnano.2c02784},
url={https://doi.org/10.1021/acsnano.2c02784}
}

@Article{Chew2023,
author={Chew, Pin Yu
and Reinhardt, Aleks},
title={Phase diagrams---Why they matter and how to predict them},
journal={J. Chem. Phys.},
year={2023},
month={Jan},
day={20},
volume={158},
number={3},
pages={030902},
issn={0021-9606},
doi={10.1063/5.0131028},
url={https://doi.org/10.1063/5.0131028}
}

@Article{Bui2024,
author={Bui, Anna T.
and Cox, Stephen J.},
title={Revisiting the {Green--Kubo} relation for friction in nanofluidics},
journal={J. Chem. Phys.},
year={2024},
month={Nov},
day={27},
volume={161},
number={20},
pages={201102},
issn={0021-9606},
doi={10.1063/5.0238363},
url={https://doi.org/10.1063/5.0238363}
}

@misc{rhodes2025,
      title={Orb-v3: atomistic simulation at scale}, 
      author={Benjamin Rhodes and Sander Vandenhaute and Vaidotas Šimkus and James Gin and Jonathan Godwin and Tim Duignan and Mark Neumann},
      year={2025},
      eprint={2504.06231},
      archivePrefix={arXiv},
      primaryClass={cond-mat.mtrl-sci},
      url={https://arxiv.org/abs/2504.06231},}

@misc{wood2025,
      title={{UMA}: A Family of Universal Models for Atoms}, 
      author={Brandon M. Wood and Misko Dzamba and Xiang Fu and Meng Gao and Muhammed Shuaibi and Luis Barroso-Luque and Kareem Abdelmaqsoud and Vahe Gharakhanyan and John R. Kitchin and Daniel S. Levine and Kyle Michel and Anuroop Sriram and Taco Cohen and Abhishek Das and Ammar Rizvi and Sushree Jagriti Sahoo and Zachary W. Ulissi and C. Lawrence Zitnick},
      year={2025},
      eprint={2506.23971},
      archivePrefix={arXiv},
      primaryClass={cs.LG},
      url={https://arxiv.org/abs/2506.23971}, 
}

@Article{Geissler2001,
author={Geissler, Phillip L.
and Dellago, Christoph
and Chandler, David
and Hutter, J{\"u}rg
and Parrinello, Michele},
title={Autoionization in Liquid Water},
journal={Science},
year={2001},
month={Mar},
day={16},
publisher={American Association for the Advancement of Science},
volume={291},
number={5511},
pages={2121-2124},
doi={10.1126/science.1056991},
url={https://doi.org/10.1126/science.1056991}
}

@Article{Caruso2026,
author={Caruso, Alessandro
and Venturin, Jacopo
and Giambagli, Lorenzo
and Rolando, Edoardo
and El-Machachi, Zakariya
and No{\'e}, Frank
and Clementi, Cecilia},
title={Extending the range of graph neural networks with global encodings},
journal={Nat. Commun.},
year={2026},
month={Feb},
day={18},
volume={17},
number={1},
pages={1855},
issn={2041-1723},
doi={10.1038/s41467-026-69715-3},
url={https://doi.org/10.1038/s41467-026-69715-3}
}

@Article{Loche2025,
author={Loche, Philip
and Huguenin-Dumittan, Kevin K.
and Honarmand, Melika
and Xu, Qianjun
and Rumiantsev, Egor
and How, Wei Bin
and Langer, Marcel F.
and Ceriotti, Michele},
title={Fast and flexible long-range models for atomistic machine learning},
journal={J. Chem. Phys.},
year={2025},
month={Apr},
day={08},
volume={162},
number={14},
pages={142501},
issn={0021-9606},
doi={10.1063/5.0251713},
url={https://doi.org/10.1063/5.0251713}
}

@Article{Kim2025,
author={Kim, Dongjin
and Wang, Xiaoyu
and Vargas, Santiago
and Zhong, Peichen
and King, Daniel S.
and Inizan, Theo Jaffrelot
and Cheng, Bingqing},
title={A Universal Augmentation Framework for Long-Range Electrostatics in Machine Learning Interatomic Potentials},
journal={J. Chem. Theory Comput.},
year={2025},
month={Dec},
day={23},
publisher={American Chemical Society},
volume={21},
number={24},
pages={12709-12724},
issn={1549-9618},
doi={10.1021/acs.jctc.5c01400},
url={https://doi.org/10.1021/acs.jctc.5c01400}
}

@Article{Niblett2021,
author={Niblett, Samuel P.
and Galib, Mirza
and Limmer, David T.},
title={Learning intermolecular forces at liquid--vapor interfaces},
journal={J. Chem. Phys.},
year={2021},
month={Oct},
day={26},
volume={155},
number={16},
pages={164101},
issn={0021-9606},
doi={10.1063/5.0067565},
url={https://doi.org/10.1063/5.0067565}
}

@Article{Yue2021,
author={Yue, Shuwen
and Muniz, Maria Carolina
and Calegari Andrade, Marcos F.
and Zhang, Linfeng
and Car, Roberto
and Panagiotopoulos, Athanassios Z.},
title={When do short-range atomistic machine-learning models fall short?},
journal={J. Chem. Phys.},
year={2021},
month={Jan},
day={21},
volume={154},
number={3},
pages={034111},
issn={0021-9606},
doi={10.1063/5.0031215},
url={https://doi.org/10.1063/5.0031215}
}

@Article{Zhang2022natcom,
author={Zhang, Chunyi
and Yue, Shuwen
and Panagiotopoulos, Athanassios Z.
and Klein, Michael L.
and Wu, Xifan},
title={Dissolving salt is not equivalent to applying a pressure on water},
journal={Nat. Commun.},
year={2022},
month={Feb},
day={10},
volume={13},
number={1},
pages={822},
issn={2041-1723},
doi={10.1038/s41467-022-28538-8},
url={https://doi.org/10.1038/s41467-022-28538-8}
}

@Article{Perkin2010,
author={Perkin, Susan
and Albrecht, Tim
and Klein, Jacob},
title={Layering and shear properties of an ionic liquid, 1-ethyl-3-methylimidazolium ethylsulfate, confined to nano-films between mica surfaces},
journal={Phys. Chem. Chem. Phys.},
year={2010},
publisher={The Royal Society of Chemistry},
volume={12},
number={6},
pages={1243-1247},
issn={1463-9076},
doi={10.1039/B920571C},
url={https://doi.org/10.1039/B920571C}
}

@misc{luise2026,
      title={Accurate and scalable exchange-correlation with deep learning}, 
      author={Giulia Luise and Chin-Wei Huang and Thijs Vogels and Derk P. Kooi and Sebastian Ehlert and Stephanie Lanius and Klaas J. H. Giesbertz and Amir Karton and Deniz Gunceler and Megan Stanley and Wessel P. Bruinsma and Lin Huang and Xinran Wei and José Garrido Torres and Abylay Katbashev and Rodrigo Chavez Zavaleta and Bálint Máté and Sékou-Oumar Kaba and Roberto Sordillo and Yingrong Chen and David B. Williams-Young and Christopher M. Bishop and Jan Hermann and Rianne van den Berg and Paola Gori-Giorgi},
      year={2026},
      eprint={2506.14665},
      archivePrefix={arXiv},
      primaryClass={physics.chem-ph},
      url={https://arxiv.org/abs/2506.14665}, 
}

@Article{Pederson2022,
author={Pederson, Ryan
and Kalita, Bhupalee
and Burke, Kieron},
title={Machine learning and density functional theory},
journal={Nat. Rev. Phys.},
year={2022},
month={Jun},
day={01},
volume={4},
number={6},
pages={357-358},
issn={2522-5820},
doi={10.1038/s42254-022-00470-2},
url={https://doi.org/10.1038/s42254-022-00470-2}
}

@Article{Kirkpatrick2021,
author={Kirkpatrick, James
and McMorrow, Brendan
and Turban, David H. P.
and Gaunt, Alexander L.
and Spencer, James S.
and Matthews, Alexander G. D. G.
and Obika, Annette
and Thiry, Louis
and Fortunato, Meire
and Pfau, David
and Castellanos, Lara Rom{\'a}n
and Petersen, Stig
and Nelson, Alexander W. R.
and Kohli, Pushmeet
and Mori-S{\'a}nchez, Paula
and Hassabis, Demis
and Cohen, Aron J.},
title={Pushing the frontiers of density functionals by solving the fractional electron problem},
journal={Science},
year={2021},
month={Dec},
day={10},
publisher={American Association for the Advancement of Science},
volume={374},
number={6573},
pages={1385-1389},
doi={10.1126/science.abj6511},
url={https://doi.org/10.1126/science.abj6511}
}

@Article{Zhang2026,
author={Zhang, Xitong
and Zhao, Hongyu
and Panter, Jack R.
and McHale, Glen
and Wells, Gary G.
and Ledesma-Aguilar, Rodrigo
and Kusumaatmaja, Halim},
title={Multiple equilibria enables tunable wetting of droplets on patterned liquid surfaces},
journal={Sci. Adv.},
year={2026},
publisher={American Association for the Advancement of Science},
volume={11},
number={38},
pages={eadw6615},
doi={10.1126/sciadv.adw6615},
url={https://doi.org/10.1126/sciadv.adw6615}
}

@Article{Liu2017,
author={Liu, Mingjie
and Wang, Shutao
and Jiang, Lei},
title={Nature-inspired superwettability systems},
journal={Nat. Rev. Mater.},
year={2017},
month={Jun},
day={27},
volume={2},
number={7},
pages={17036},
issn={2058-8437},
doi={10.1038/natrevmats.2017.36},
url={https://doi.org/10.1038/natrevmats.2017.36}
}

@Article{Han2010,
author={Han, Sungho
and Choi, M. Y.
and Kumar, Pradeep
and Stanley, H. Eugene},
title={Phase transitions in confined water nanofilms},
journal={Nat. Phys.},
year={2010},
month={Sep},
day={01},
volume={6},
number={9},
pages={685-689},
issn={1745-2481},
doi={10.1038/nphys1708},
url={https://doi.org/10.1038/nphys1708}
}

@Article{Zimmermann2024,
author={Zimmermann, Toni
and Samm{\"u}ller, Florian
and Hermann, Sophie
and Schmidt, Matthias
and de las Heras, Daniel},
title={Neural force functional for non-equilibrium many-body colloidal systems},
journal={Mach. Learn.: Sci. Technol.},
year={2024},
month={Sep},
day={02},
publisher={IOP Publishing},
volume={5},
number={3},
pages={035062},
issn={2632-2153},
doi={10.1088/2632-2153/ad7191},
url={https://doi.org/10.1088/2632-2153/ad7191}
}

@Article{Zou2021,
author={Zou, An
and Poudel, Sajag
and Gupta, Manish
and Maroo, Shalabh C.},
title={Disjoining Pressure of Water in Nanochannels},
journal={Nano Lett},
year={2021},
month={Sep},
day={22},
publisher={American Chemical Society},
volume={21},
number={18},
pages={7769-7774},
issn={1530-6984},
doi={10.1021/acs.nanolett.1c02726},
url={https://doi.org/10.1021/acs.nanolett.1c02726}
}

@article{Glitsch2025,
  title = {Neural density functional theory in higher dimensions with convolutional layers},
  author = {Glitsch, Felix and Weimar, Jens and Oettel, Martin},
  journal = {Phys. Rev. E},
  volume = {111},
  issue = {5},
  pages = {055305},
  numpages = {9},
  year = {2025},
  month = {May},
  publisher = {American Physical Society},
  doi = {10.1103/PhysRevE.111.055305},
  url = {https://link.aps.org/doi/10.1103/PhysRevE.111.055305}
}

@Article{Cockrell2021,
author={Cockrell, C.
and Brazhkin, V. V.
and Trachenko, K.},
title={Transition in the supercritical state of matter: Review of experimental evidence},
journal={Phys. Rep},
year={2021},
month={Dec},
day={20},
volume={941},
pages={1-27},
issn={0370-1573},
url={https://www.sciencedirect.com/science/article/pii/S0370157321003732}
}

@Article{Pan2016,
author={Pan, Ding
and Galli, Giulia},
title={The fate of carbon dioxide in water-rich fluids under extreme conditions},
journal={Sci. Adv.},
year={2016},
publisher={American Association for the Advancement of Science},
volume={2},
number={10},
pages={e1601278},
doi={10.1126/sciadv.1601278},
url={https://doi.org/10.1126/sciadv.1601278}
}

@Article{Pan2013,
author={Pan, Ding
and Spanu, Leonardo
and Harrison, Brandon
and Sverjensky, Dimitri A.
and Galli, Giulia},
title={Dielectric properties of water under extreme conditions and transport of carbonates in the deep Earth},
journal={Proc. Natl. Acad. Sci. U.S.A.},
year={2013},
month={Apr},
day={23},
publisher={Proc. Natl. Acad. Sci. U.S.A},
volume={110},
number={17},
pages={6646-6650},
doi={10.1073/pnas.1221581110},
url={https://doi.org/10.1073/pnas.1221581110}
}

@article{Vega1995,
  title = {Location of the {Fisher-Widom} line for systems interacting through short-ranged potentials},
  author = {Vega, C. and Rull, L. F. and Lago, S.},
  journal = {Phys. Rev. E},
  volume = {51},
  issue = {4},
  pages = {3146--3155},
  numpages = {0},
  year = {1995},
  month = {Apr},
  publisher = {American Physical Society},
  doi = {10.1103/PhysRevE.51.3146},
  url = {https://link.aps.org/doi/10.1103/PhysRevE.51.3146}
}

@Article{Li2024,
author={Li, Xinyang
and Jin, Yuliang},
title={Thermodynamic crossovers in supercritical fluids},
journal={Proc. Natl. Acad. Sci. U.S.A},
year={2024},
month={Apr},
day={30},
publisher={Proc. Natl. Acad. Sci. U.S.A},
volume={121},
number={18},
pages={e2400313121},
doi={10.1073/pnas.2400313121},
url={https://doi.org/10.1073/pnas.2400313121}
}

@Article{Remsing2016,
author={Remsing, Richard C.
and Liu, Shule
and Weeks, John D.},
title={Long-ranged contributions to solvation free energies from theory and short-ranged models},
journal={Proc. Natl. Acad. Sci. U.S.A},
year={2016},
month={Mar},
day={15},
publisher={Proc. Natl. Acad. Sci. U.S.A},
volume={113},
number={11},
pages={2826},
doi={10.1073/pnas.1521570113},
url={https://doi.org/10.1073/pnas.1521570113}
}

@Article{Cox2020,
author={Cox, Stephen J.},
title={Dielectric response with short-ranged electrostatics},
journal={Proc. Natl. Acad. Sci. U.S.A},
year={2020},
month={Aug},
day={18},
publisher={Proc. Natl. Acad. Sci. U.S.A},
volume={117},
number={33},
pages={19746-19752},
doi={10.1073/pnas.2005847117},
url={https://doi.org/10.1073/pnas.2005847117}
}

@Article{Ko2021,
author={Ko, Tsz Wai
and Finkler, Jonas A.
and Goedecker, Stefan
and Behler, J{\"o}rg},
title={A fourth-generation high-dimensional neural network potential with accurate electrostatics including non-local charge transfer},
journal={Nat. Commun},
year={2021},
month={Jan},
day={15},
volume={12},
number={1},
pages={398},
issn={2041-1723},
doi={10.1038/s41467-020-20427-2},
url={https://doi.org/10.1038/s41467-020-20427-2}
}

@Article{Zhang2022,
author={Zhang, Linfeng
and Wang, Han
and Muniz, Maria Carolina
and Panagiotopoulos, Athanassios Z.
and Car, Roberto
and E, Weinan},
title={A deep potential model with long-range electrostatic interactions},
journal={J. Chem. Phys},
year={2022},
month={Mar},
day={24},
volume={156},
number={12},
pages={124107},
issn={0021-9606},
doi={10.1063/5.0083669},
url={https://doi.org/10.1063/5.0083669}
}

@Article{Cheng2025,
author={Cheng, Bingqing},
title={Latent {Ewald} summation for machine learning of long-range interactions},
journal={npj Comput. Mater.},
year={2025},
month={Mar},
day={26},
volume={11},
number={1},
pages={80},
issn={2057-3960},
doi={10.1038/s41524-025-01577-7},
url={https://doi.org/10.1038/s41524-025-01577-7}
}

@Article{Rodgers2008,
author={Rodgers, Jocelyn M.
and Weeks, John D.},
title={Local molecular field theory for the treatment of electrostatics},
journal={J. Phys.: Condens. Matter},
year={2008},
month={Nov},
day={12},
volume={20},
number={49},
pages={494206},
issn={0953-8984},
doi={10.1088/0953-8984/20/49/494206},
url={https://doi.org/10.1088/0953-8984/20/49/494206}
}

@Article{Gao2022,
author={Gao, Ang
and Remsing, Richard C.},
title={Self-consistent determination of long-range electrostatics in neural network potentials},
journal={Nat. Commun},
year={2022},
month={Mar},
day={23},
volume={13},
number={1},
pages={1572},
issn={2041-1723},
doi={10.1038/s41467-022-29243-2},
url={https://doi.org/10.1038/s41467-022-29243-2}
}

@Article{White2021,
author={White, Martin T.
and Bianchi, Giuseppe
and Chai, Lei
and Tassou, Savvas A.
and Sayma, Abdulnaser I.},
title={Review of supercritical {CO$_2$} technologies and systems for power generation},
journal={Appl. Therm. Eng},
year={2021},
month={Feb},
day={25},
volume={185},
pages={116447},
issn={1359-4311},
url={https://www.sciencedirect.com/science/article/pii/S1359431120339235}
}

@Article{McMillan2010,
author={McMillan, Paul F.
and Stanley, H. Eugene},
title={Going supercritical},
journal={Nat. Phys.},
year={2010},
month={Jul},
day={01},
volume={6},
number={7},
pages={479-480},
issn={1745-2481},
doi={10.1038/nphys1711},
url={https://doi.org/10.1038/nphys1711}
}

@Article{Gallo2014,
author={Gallo, P.
and Corradini, D.
and Rovere, M.},
title={Widom line and dynamical crossovers as routes to understand supercritical water},
journal={Nat. Commun},
year={2014},
month={Dec},
day={16},
volume={5},
number={1},
pages={5806},
issn={2041-1723},
doi={10.1038/ncomms6806},
url={https://doi.org/10.1038/ncomms6806}
}

@Article{Fomin2015,
author={Fomin, Yu. D.
and Ryzhov, V. N.
and Tsiok, E. N.
and Brazhkin, V. V.},
title={Thermodynamic properties of supercritical carbon dioxide: {Widom} and {Frenkel} lines},
journal={Phys. Rev. E},
year={2015},
month={Feb},
day={09},
publisher={American Physical Society},
volume={91},
number={2},
pages={022111},
doi={10.1103/PhysRevE.91.022111},
url={https://doi.org/10.1103/PhysRevE.91.022111}
}

@article{Kohn1999,
  title = {Nobel Lecture: Electronic structure of matter---wave functions and density functionals},
  author = {Kohn, W.},
  journal = {Rev. Mod. Phys.},
  volume = {71},
  issue = {5},
  pages = {1253--1266},
  numpages = {0},
  year = {1999},
  month = {Oct},
  publisher = {American Physical Society},
  doi = {10.1103/RevModPhys.71.1253},
  url = {https://link.aps.org/doi/10.1103/RevModPhys.71.1253}
}

@Article{Evans1979,
author={Evans, R.},
title={The nature of the liquid-vapour interface and other topics in the statistical mechanics of non-uniform, classical fluids},
journal={Adv. Phys},
year={1979},
month={Apr},
day={01},
publisher={Taylor {\&} Francis},
volume={28},
number={2},
pages={143-200},
issn={0001-8732},
doi={10.1080/00018737900101365},
url={https://doi.org/10.1080/00018737900101365}
}

@article{Gorelli2006,
  title = {Liquidlike Behavior of Supercritical Fluids},
  author = {Gorelli, F. and Santoro, M. and Scopigno, T. and Krisch, M. and Ruocco, G.},
  journal = {Phys. Rev. Lett.},
  volume = {97},
  issue = {24},
  pages = {245702},
  numpages = {4},
  year = {2006},
  month = {Dec},
  publisher = {American Physical Society},
  doi = {10.1103/PhysRevLett.97.245702},
  url = {https://link.aps.org/doi/10.1103/PhysRevLett.97.245702}
}

@article{Pipich2018,
  title = {Densification of Supercritical Carbon Dioxide Accompanied by Droplet Formation When Passing the {Widom} Line},
  author = {Pipich, Vitaliy and Schwahn, Dietmar},
  journal = {Phys. Rev. Lett.},
  volume = {120},
  issue = {14},
  pages = {145701},
  numpages = {5},
  year = {2018},
  month = {Apr},
  publisher = {American Physical Society},
  doi = {10.1103/PhysRevLett.120.145701},
  url = {https://link.aps.org/doi/10.1103/PhysRevLett.120.145701}
}

@article{Ji2025,
  title = {Machine-Learning Interatomic Potentials for Long-Range Systems},
  author = {Ji, Yajie and Liang, Jiuyang and Xu, Zhenli},
  journal = {Phys. Rev. Lett.},
  volume = {135},
  issue = {17},
  pages = {178001},
  numpages = {8},
  year = {2025},
  month = {Oct},
  publisher = {American Physical Society},
  doi = {10.1103/ssp9-7s81},
  url = {https://link.aps.org/doi/10.1103/ssp9-7s81}
}

@Article{Fisher1969,
author={Fisher, Michael E.
and Widom, B.},
title={Decay of Correlations in Linear Systems},
journal={J. Chem. Phys},
year={1969},
month={May},
day={01},
volume={50},
number={9},
pages={3756-3772},
issn={0021-9606},
doi={10.1063/1.1671624},
url={https://doi.org/10.1063/1.1671624}
}

@Article{Xu2005,
author={Xu, Limei
and Kumar, Pradeep
and Buldyrev, S. V.
and Chen, S.-H.
and Poole, P. H.
and Sciortino, F.
and Stanley, H. E.},
title={Relation between the {Widom} line and the dynamic crossover in systems with a liquid--liquid phase transition},
journal={Proc. Natl. Acad. Sci. U.S.A},
year={2005},
month={Nov},
day={15},
publisher={Proc. Natl. Acad. Sci. U.S.A},
volume={102},
number={46},
pages={16558-16562},
doi={10.1073/pnas.0507870102},
url={https://doi.org/10.1073/pnas.0507870102}
}

@Article{Hsiao2025,
author={Hsiao, Kuang-Yuan
and Chung, Ren-Jei
and Chen, Chin-Fu
and Chang, Pi-Pai
and Tsai, Teh-Hua},
title={Review on Supercritical Carbon Dioxide in Energy Storage Systems: Advances and Outlook},
journal={Energy Fuels},
year={2025},
month={Jul},
day={03},
publisher={American Chemical Society},
volume={39},
number={26},
pages={12289-12308},
issn={0887-0624},
doi={10.1021/acs.energyfuels.5c00825},
url={https://doi.org/10.1021/acs.energyfuels.5c00825}
}

@Article{Dijkstra2000,
author={Dijkstra, Marjolein
and Evans, Robert},
title={A simulation study of the decay of the pair correlation function in simple fluids},
journal={J. Chem. Phys},
year={2000},
month={Jan},
day={15},
volume={112},
number={3},
pages={1449-1456},
issn={0021-9606},
doi={10.1063/1.480598},
url={https://doi.org/10.1063/1.480598}
}

@Article{Simeoni2010,
author={Simeoni, G. G.
and Bryk, T.
and Gorelli, F. A.
and Krisch, M.
and Ruocco, G.
and Santoro, M.
and Scopigno, T.},
title={The {Widom} line as the crossover between liquid-like and gas-like behaviour in supercritical fluids},
journal={Nat. Phys.},
year={2010},
month={Jul},
day={01},
volume={6},
number={7},
pages={503-507},
issn={1745-2481},
doi={10.1038/nphys1683},
url={https://doi.org/10.1038/nphys1683}
}

@Article{Span1996,
author={Span, Roland
and Wagner, Wolfgang},
title={A New Equation of State for Carbon Dioxide Covering the Fluid Region from the Triple‐Point Temperature to 1100$\,${K} at Pressures up to 800$\,${MPa}},
journal={J. Phys. Chem. Ref. Data},
year={1996},
month={Nov},
day={01},
volume={25},
number={6},
pages={1509-1596},
issn={0047-2689},
doi={10.1063/1.555991},
url={https://doi.org/10.1063/1.555991}
}

@Article{Yang2025,
author={Yang, Jinni
and Pan, Runtong
and Sun, Jikai
and Wu, Jianzhong},
title={High-Dimensional Operator Learning for Molecular Density Functional Theory},
journal={J. Chem. Theory Comput},
year={2025},
month={Jun},
day={24},
publisher={American Chemical Society},
volume={21},
number={12},
pages={5905-5915},
issn={1549-9618},
doi={10.1021/acs.jctc.5c00484},
url={https://doi.org/10.1021/acs.jctc.5c00484}
}

@misc{nist_webbook,
  author       = {P. J. Linstrom and W. G. Mallard},
  title        = {{NIST Chemistry WebBook}, {NIST Standard Reference Database Number} 69},
  howpublished = {\url{https://doi.org/10.18434/T4D303}},
  note         = {{National Institute of Standards and Technology}, Gaithersburg MD, 20899. Retrieved October 29, 2025},
  year         = {2025},
  editor       = {P. J. Linstrom and W. G. Mallard},
}

@book{ProctorBook,
  title={The Liquid and Supercritical Fluid States of Matter},
  author={Proctor, J.E.},
  isbn={9780429957925},
  year={2020},
  publisher={CRC Press}
}

@Article{Kapil2022,
author={Kapil, Venkat
and Schran, Christoph
and Zen, Andrea
and Chen, Ji
and Pickard, Chris J.
and Michaelides, Angelos},
title={The first-principles phase diagram of monolayer nanoconfined water},
journal={Nature},
year={2022},
month={Sep},
day={01},
volume={609},
number={7927},
pages={512-516},
issn={1476-4687},
doi={10.1038/s41586-022-05036-x},
url={https://doi.org/10.1038/s41586-022-05036-x}
}

@Article{Zhang2025,
author={Zhang, Chunyi
and Yu, Zheng
and Car, Roberto
and Selloni, Annabella},
title={Tuning water dissociation at oxide--electrolyte interfaces with electric fields},
journal={Proc. Natl. Acad. Sci. U.S.A},
year={2025},
month={Aug},
day={26},
publisher={Proc. Natl. Acad. Sci. U.S.A},
volume={122},
number={34},
pages={e2505929122},
doi={10.1073/pnas.2505929122},
url={https://doi.org/10.1073/pnas.2505929122}
}

@Article{Reinhardt2021,
author={Reinhardt, Aleks
and Cheng, Bingqing},
title={Quantum-mechanical exploration of the phase diagram of water},
journal={Nat. Commun},
year={2021},
month={Jan},
day={26},
volume={12},
number={1},
pages={588},
issn={2041-1723},
doi={10.1038/s41467-020-20821-w},
url={https://doi.org/10.1038/s41467-020-20821-w}
}

@Article{Abascal2005,
author={Abascal, J. L. F.
and Vega, C.},
title={A general purpose model for the condensed phases of water: {TIP4P}/2005},
journal={J. Chem. Phys},
year={2005},
month={Dec},
day={19},
volume={123},
number={23},
pages={234505},
issn={0021-9606},
doi={10.1063/1.2121687},
url={https://doi.org/10.1063/1.2121687}
}

@Article{Potoff2001,
author={Potoff, Jeffrey J.
and Siepmann, J. Ilja},
title={Vapor--liquid equilibria of mixtures containing alkanes, carbon dioxide, and nitrogen},
journal={AIChE J.},
year={2001},
month={Jul},
day={01},
publisher={John Wiley {\&} Sons, Ltd},
volume={47},
number={7},
pages={1676-1682},
issn={0001-1541},
doi={10.1002/aic.690470719},
url={https://doi.org/10.1002/aic.690470719}
}

@Article{Vega2006,
author={Vega, C.
and Abascal, J. L. F.
and Nezbeda, I.},
title={Vapor-liquid equilibria from the triple point up to the critical point for the new generation of {TIP4P}-like models: {TIP4P/Ew}, {TIP4P/2005}, and {TIP4P/ice}},
journal={J. Chem. Phys},
year={2006},
month={Jul},
day={18},
volume={125},
number={3},
pages={034503},
issn={0021-9606},
doi={10.1063/1.2215612},
url={https://doi.org/10.1063/1.2215612}
}

@Article{Wohlfahrt2020,
author={Wohlfahrt, Oliver
and Dellago, Christoph
and Sega, Marcello},
title={Ab initio structure and thermodynamics of the {RPBE-D3} water/vapor interface by neural-network molecular dynamics},
journal={J. Chem. Phys},
year={2020},
month={Oct},
day={14},
volume={153},
number={14},
pages={144710},
issn={0021-9606},
doi={10.1063/5.0021852},
url={https://doi.org/10.1063/5.0021852}
}

@Article{Sanchez-Burgos2023,
author={Sanchez-Burgos, Ignacio
and Muniz, Maria Carolina
and Espinosa, Jorge R.
and Panagiotopoulos, Athanassios Z.},
title={A Deep Potential model for liquid--vapor equilibrium and cavitation rates of water},
journal={J. Chem. Phys},
year={2023},
month={May},
day={09},
volume={158},
number={18},
pages={184504},
issn={0021-9606},
doi={10.1063/5.0144500},
url={https://doi.org/10.1063/5.0144500}
}

@Article{Goeminne2025,
author={Goeminne, Ruben
and Van Speybroeck, Veronique},
title={Ab Initio Predictions of Adsorption in Flexible Metal--Organic Frameworks for Water Harvesting Applications},
journal={J. Am. Chem. Soc.},
year={2025},
month={Jan},
day={29},
publisher={American Chemical Society},
volume={147},
number={4},
pages={3615-3630},
issn={0002-7863},
doi={10.1021/jacs.4c15287},
url={https://doi.org/10.1021/jacs.4c15287}
}

@Article{Radhakrishnan2021,
author={Radhakrishnan, Ravi},
title={A survey of multiscale modeling: Foundations, historical milestones, current status, and future prospects},
journal={AIChE J.},
year={2021},
month={Mar},
day={01},
publisher={John Wiley {\&} Sons, Ltd},
volume={67},
number={3},
pages={e17026},
issn={0001-1541},
doi={10.1002/aic.17026},
url={https://doi.org/10.1002/aic.17026}
}

@Article{Menon2024,
author={Menon, Sarath
and Lysogorskiy, Yury
and Knoll, Alexander L. M.
and Leimeroth, Niklas
and Poul, Marvin
and Qamar, Minaam
and Janssen, Jan
and Mrovec, Matous
and Rohrer, Jochen
and Albe, Karsten
and Behler, J{\"o}rg
and Drautz, Ralf
and Neugebauer, J{\"o}rg},
title={From electrons to phase diagrams with machine learning potentials using pyiron based automated workflows},
journal={npj Comput. Mater.},
year={2024},
month={Nov},
day={17},
volume={10},
number={1},
pages={261},
issn={2057-3960},
doi={10.1038/s41524-024-01441-0},
url={https://doi.org/10.1038/s41524-024-01441-0}
}

@article{Rosenfeld1989,
  title = {Free-energy model for the inhomogeneous hard-sphere fluid mixture and density-functional theory of freezing},
  author = {Rosenfeld, Yaakov},
  journal = {Phys. Rev. Lett.},
  volume = {63},
  issue = {9},
  pages = {980--983},
  numpages = {0},
  year = {1989},
  month = {Aug},
  publisher = {American Physical Society},
  doi = {10.1103/PhysRevLett.63.980},
  url = {https://link.aps.org/doi/10.1103/PhysRevLett.63.980}
}

@Article{Roth2002,
author={Roth, R.
and Evans, R.
and Lang, A.
and Kahl, G.},
title={Fundamental measure theory for hard-sphere mixtures revisited: the {W}hite {B}ear version},
journal={J. Phys. Condens. Matter},
year={2002},
month={Nov},
day={08},
volume={14},
number={46},
pages={12063},
issn={0953-8984},
doi={10.1088/0953-8984/14/46/313},
url={https://doi.org/10.1088/0953-8984/14/46/313}
}

@Article{Miller2002,
author={Miller III, T. F.
and Eleftheriou, M.
and Pattnaik, P.
and Ndirango, A.
and Newns, D.
and Martyna, G. J.},
title={Symplectic quaternion scheme for biophysical molecular dynamics},
journal={J. Chem. Phys.},
year={2002},
month={May},
day={22},
volume={116},
number={20},
pages={8649-8659},
issn={0021-9606},
doi={10.1063/1.1473654},
url={https://doi.org/10.1063/1.1473654}
}

@Article{deCarvalho1994,
author={de Carvalho, R. J. F. Leote
and Evans, R.
and Hoyle, D. C.
and Henderson, J. R.},
title={The decay of the pair correlation function in simple fluids: long- versus short-ranged potentials},
journal={J. Phys.: Condens. Matter},
year={1994},
month={Oct},
day={31},
volume={6},
number={44},
pages={9275},
issn={0953-8984},
doi={10.1088/0953-8984/6/44/008},
url={https://doi.org/10.1088/0953-8984/6/44/008}
}

@misc{Batatia2026,
      title={{MACE-POLAR}-1: A Polarisable Electrostatic Foundation Model for Molecular Chemistry}, 
      author={Ilyes Batatia and William J. Baldwin and Domantas Kuryla and Joseph Hart and Elliott Kasoar and Alin M. Elena and Harry Moore and Mikołaj J. Gawkowski and Benjamin X. Shi and Venkat Kapil and Panagiotis Kourtis and Ioan-Bogdan Magd\u{a}u and G{\'a}bor Cs{\'a}nyi},
      year={2026},
      eprint={2602.19411},
      archivePrefix={arXiv},
      primaryClass={physics.chem-ph},
      url={https://arxiv.org/abs/2602.19411}, 
}

@Article{Takaiwa2008,
author={Takaiwa, Daisuke
and Hatano, Itaru
and Koga, Kenichiro
and Tanaka, Hideki},
title={Phase diagram of water in carbon nanotubes},
journal={Proc. Natl. Acad. Sci. U.S.A},
year={2008},
month={Jan},
day={08},
publisher={Proc. Natl. Acad. Sci. U.S.A},
volume={105},
number={1},
pages={39-43},
doi={10.1073/pnas.0707917105},
url={https://doi.org/10.1073/pnas.0707917105}
}

@Article{Fumagalli2018,
author={Fumagalli, L.
and Esfandiar, A.
and Fabregas, R.
and Hu, S.
and Ares, P.
and Janardanan, A.
and Yang, Q.
and Radha, B.
and Taniguchi, T.
and Watanabe, K.
and Gomila, G.
and Novoselov, K. S.
and Geim, A. K.},
title={Anomalously low dielectric constant of confined water},
journal={Science},
year={2018},
month={Jun},
day={22},
publisher={American Association for the Advancement of Science},
volume={360},
number={6395},
pages={1339-1342},
doi={10.1126/science.aat4191},
url={https://doi.org/10.1126/science.aat4191}
}

@Article{Wang2025,
author={Wang, R.
and Souilamas, M.
and Esfandiar, A.
and Fabregas, R.
and Benaglia, S.
and Nevison-Andrews, H.
and Yang, Q.
and Normansell, J.
and Ares, P.
and Ferrari, G.
and Principi, A.
and Geim, A. K.
and Fumagalli, L.},
title={In-plane dielectric constant and conductivity of confined water},
journal={Nature},
year={2025},
month={Oct},
day={01},
volume={646},
number={8085},
pages={606-610},
issn={1476-4687},
doi={10.1038/s41586-025-09558-y},
url={https://doi.org/10.1038/s41586-025-09558-y}
}

@Article{Radha2016,
author={Radha, B.
and Esfandiar, A.
and Wang, F. C.
and Rooney, A. P.
and Gopinadhan, K.
and Keerthi, A.
and Mishchenko, A.
and Janardanan, A.
and Blake, P.
and Fumagalli, L.
and Lozada-Hidalgo, M.
and Garaj, S.
and Haigh, S. J.
and Grigorieva, I. V.
and Wu, H. A.
and Geim, A. K.},
title={Molecular transport through capillaries made with atomic-scale precision},
journal={Nature},
year={2016},
month={Oct},
day={01},
volume={538},
number={7624},
pages={222-225},
issn={1476-4687},
doi={10.1038/nature19363},
url={https://doi.org/10.1038/nature19363}
}

@article{Tang2004,
  title = {Modeling inhomogeneous van der Waals fluids using an analytical direct correlation function},
  author = {Tang, Yiping and Wu, Jianzhong},
  journal = {Phys. Rev. E},
  volume = {70},
  issue = {1},
  pages = {011201},
  numpages = {8},
  year = {2004},
  month = {Jul},
  publisher = {American Physical Society},
  doi = {10.1103/PhysRevE.70.011201},
  url = {https://link.aps.org/doi/10.1103/PhysRevE.70.011201}
}

@Article{Archer2017,
author={Archer, Andrew J.
and Chacko, Blesson
and Evans, Robert},
title={The standard mean-field treatment of inter-particle attraction in classical DFT is better than one might expect},
journal={J. Chem. Phys.},
year={2017},
month={Jul},
day={17},
volume={147},
number={3},
pages={034501},
issn={0021-9606},
doi={10.1063/1.4993175},
url={https://doi.org/10.1063/1.4993175}
}

@Article{Simon2025,
author={Simon, Alessandro
and Belloni, Luc
and Borgis, Daniel
and Oettel, Martin},
title={The orientational structure of a model patchy particle fluid: Simulations, integral equations, density functional theory, and machine learning},
journal={J. Chem. Phys.},
year={2025},
month={Jan},
day={16},
volume={162},
number={3},
pages={034503},
issn={0021-9606},
doi={10.1063/5.0248694},
url={https://doi.org/10.1063/5.0248694}
}

@article{Dijkman2025,
  title = {Learning Neural Free-Energy Functionals with Pair-Correlation Matching},
  author = {Dijkman, Jacobus and Dijkstra, Marjolein and van Roij, Ren\'e and Welling, Max and van de Meent, Jan-Willem and Ensing, Bernd},
  journal = {Phys. Rev. Lett.},
  volume = {134},
  issue = {5},
  pages = {056103},
  numpages = {7},
  year = {2025},
  month = {Feb},
  publisher = {American Physical Society},
  doi = {10.1103/PhysRevLett.134.056103},
  url = {https://link.aps.org/doi/10.1103/PhysRevLett.134.056103}
}

@Article{Jeanmairet2013,
author={Jeanmairet, Guillaume
and Levesque, Maximilien
and Vuilleumier, Rodolphe
and Borgis, Daniel},
title={Molecular Density Functional Theory of Water},
journal={J. Phys. Chem. Lett.},
year={2013},
month={Feb},
day={21},
publisher={American Chemical Society},
volume={4},
number={4},
pages={619-624},
doi={10.1021/jz301956b},
url={https://doi.org/10.1021/jz301956b}
}

@Article{Jeanmairet2026,
author={Jeanmairet, Guillaume
and Belloni, Luc
and Borgis, Daniel},
title={A molecular density functional theory of aqueous electrolytic solution},
journal={J. Chem. Phys.},
year={2026},
month={Mar},
day={09},
volume={164},
number={10},
pages={104104},
issn={0021-9606},
doi={10.1063/5.0311853},
url={https://doi.org/10.1063/5.0311853}
}

@Article{Mazitov2025,
author={Mazitov, Arslan
and Bigi, Filippo
and Kellner, Matthias
and Pegolo, Paolo
and Tisi, Davide
and Fraux, Guillaume
and Pozdnyakov, Sergey
and Loche, Philip
and Ceriotti, Michele},
title={PET-MAD as a lightweight universal interatomic potential for advanced materials modeling},
journal={Nat. Commun.},
year={2025},
month={Nov},
day={27},
volume={16},
number={1},
pages={10653},
issn={2041-1723},
doi={10.1038/s41467-025-65662-7},
url={https://doi.org/10.1038/s41467-025-65662-7}
}

@Inbook{Simon2026,
author={Simon, Alessandro
and Oettel, Martin},
editor={te Vrugt, Michael},
title={Machine Learning Approaches to Classical Density Functional Theory},
bookTitle={Artificial Intelligence and Intelligent Matter: Nanoscience, Soft Matter, Philosophy},
year={2026},
publisher={Springer Nature Switzerland},
address={Cham},
pages={83-113},
isbn={978-3-032-04129-6},
doi={10.1007/978-3-032-04129-6_6},
url={https://doi.org/10.1007/978-3-032-04129-6_6}
}

@Article{Roth2010,
author={Roth, Roland},
title={Fundamental measure theory for hard-sphere mixtures: a review},
journal={J. Phys. Condens. Matter},
year={2010},
month={Jan},
day={27},
volume={22},
number={6},
pages={063102},
issn={0953-8984},
doi={10.1088/0953-8984/22/6/063102},
url={https://doi.org/10.1088/0953-8984/22/6/063102}
}

@Article{Zong2020,
author={Zong, Hongxiang
and Wiebe, Heather
and Ackland, Graeme J.},
title={Understanding high pressure molecular hydrogen with a hierarchical machine-learned potential},
journal={Nat. Commun.},
year={2020},
month={Oct},
day={06},
volume={11},
number={1},
pages={5014},
issn={2041-1723},
doi={10.1038/s41467-020-18788-9},
url={https://doi.org/10.1038/s41467-020-18788-9}
}

@Article{Magdau2023,
author={Magd{\u{a}}u, Ioan-Bogdan
and Arismendi-Arrieta, Daniel J.
and Smith, Holly E.
and Grey, Clare P.
and Hermansson, Kersti
and Cs{\'a}nyi, G{\'a}bor},
title={Machine learning force fields for molecular liquids: Ethylene Carbonate/Ethyl Methyl Carbonate binary solvent},
journal={npj Comput. Mater.},
year={2023},
month={Aug},
day={17},
volume={9},
number={1},
pages={146},
issn={2057-3960},
doi={10.1038/s41524-023-01100-w},
url={https://doi.org/10.1038/s41524-023-01100-w}
}

@Article{Wang2018,
author={Wang, Han
and Zhang, Linfeng
and Han, Jiequn
and E, Weinan},
title={{DeePMD}-kit: A deep learning package for many-body potential energy representation and molecular dynamics},
journal={Comput. Phys. Commun},
year={2018},
month={Jul},
day={01},
volume={228},
pages={178-184},
issn={0010-4655},
url={https://www.sciencedirect.com/science/article/pii/S0010465518300882}
}

@Article{Mathur2023,
author={Mathur, Reha
and Muniz, Maria Carolina
and Yue, Shuwen
and Car, Roberto
and Panagiotopoulos, Athanassios Z.},
title={First-Principles-Based Machine Learning Models for Phase Behavior and Transport Properties of {CO}$_2$},
journal={J. Phys. Chem. B},
year={2023},
month={May},
day={25},
publisher={American Chemical Society},
volume={127},
number={20},
pages={4562-4569},
issn={1520-6106},
doi={10.1021/acs.jpcb.3c00610},
url={https://doi.org/10.1021/acs.jpcb.3c00610}
}

@article{Peng2016,
  title = {Versatile van der {Waals} Density Functional Based on a Meta-Generalized Gradient Approximation},
  author = {Peng, Haowei and Yang, Zeng-Hui and Perdew, John P. and Sun, Jianwei},
  journal = {Phys. Rev. X},
  volume = {6},
  issue = {4},
  pages = {041005},
  numpages = {15},
  year = {2016},
  month = {Oct},
  publisher = {American Physical Society},
  doi = {10.1103/PhysRevX.6.041005},
  url = {https://link.aps.org/doi/10.1103/PhysRevX.6.041005}
}

@article{Becke1988,
  title = {Density-functional exchange-energy approximation with correct asymptotic behavior},
  author = {Becke, A. D.},
  journal = {Phys. Rev. A},
  volume = {38},
  issue = {6},
  pages = {3098--3100},
  numpages = {0},
  year = {1988},
  month = {Sep},
  publisher = {American Physical Society},
  doi = {10.1103/PhysRevA.38.3098},
  url = {https://link.aps.org/doi/10.1103/PhysRevA.38.3098}
}

@Article{Grimme2010,
author={Grimme, Stefan
and Antony, Jens
and Ehrlich, Stephan
and Krieg, Helge},
title={A consistent and accurate ab initio parametrization of density functional dispersion correction ({DFT-D}) for the 94 elements {H-Pu}},
journal={J. Chem. Phys},
year={2010},
month={Apr},
day={16},
volume={132},
number={15},
pages={154104},
issn={0021-9606},
doi={10.1063/1.3382344},
url={https://doi.org/10.1063/1.3382344}
}

@Article{bui2025dielectro,
author={Bui, Anna T.
and Cox, Stephen J.},
title={Dielectrocapillarity for exquisite control of fluids},
journal={Nat. Commun.},
year={2026},
month={Feb},
day={12},
volume={17},
number={1},
pages={2661},
issn={2041-1723},
doi={10.1038/s41467-026-69482-1},
url={https://doi.org/10.1038/s41467-026-69482-1}
}

@Article{zhou2026,
author={Zhou, Katie L. Y.
and Bui, Anna T.
and Cox, Stephen J.},
title={Roles of Bulk and Surface Thermodynamics in the Selective Adsorption of a Confined Azeotropic Mixture},
journal={J. Phys. Chem. B},
year={2026},
month={Apr},
day={23},
publisher={American Chemical Society},
volume={130},
number={16},
pages={4455-4466},
issn={1520-6106},
doi={10.1021/acs.jpcb.6c00640},
url={https://doi.org/10.1021/acs.jpcb.6c00640}
}

@article{Perdew1996,
  title = {Generalized Gradient Approximation Made Simple},
  author = {Perdew, John P. and Burke, Kieron and Ernzerhof, Matthias},
  journal = {Phys. Rev. Lett.},
  volume = {77},
  issue = {18},
  pages = {3865--3868},
  numpages = {0},
  year = {1996},
  month = {Oct},
  publisher = {American Physical Society},
  doi = {10.1103/PhysRevLett.77.3865},
  url = {https://link.aps.org/doi/10.1103/PhysRevLett.77.3865}
}

@Article{Yang2020,
author={Yang, Qian
and Sun, P. Z.
and Fumagalli, L.
and Stebunov, Y. V.
and Haigh, S. J.
and Zhou, Z. W.
and Grigorieva, I. V.
and Wang, F. C.
and Geim, A. K.},
title={Capillary condensation under atomic-scale confinement},
journal={Nature},
year={2020},
month={Dec},
day={01},
volume={588},
number={7837},
pages={250-253},
issn={1476-4687},
doi={10.1038/s41586-020-2978-1},
url={https://doi.org/10.1038/s41586-020-2978-1}
}

@article{yeh1999ewald,
  title={Ewald summation for systems with slab geometry},
  author={Yeh, In-Chul and Berkowitz, Max L},
  journal={J. Chem. Phys.},
  volume={111},
  number={7},
  pages={3155--3162},
  year={1999},
  publisher={American Institute of Physics}
}

@article{parker2026false,
  title={False Metallization in Short-Ranged Machine Learned
                  Interatomic Potentials},
  author={Parker, Isaac J and Hoffmann, Mandy J and Baldwin, William J
                  and Han, Shuang and Gupta, Srishti and Fong, Kara D
                  and Michaelides, Angelos and Schran, Christoph and
                  De, Sandip and Cs{\'a}nyi, G{\'a}bor},
  journal={arXiv preprint arXiv:2603.04228},
  year={2026},
  doi={https://doi.org/10.48550/arXiv.2603.04228}
}

@article{chen2016two,
  title={Two dimensional ice from first principles: Structures and
                  phase transitions},
  author={Chen, Ji and Schusteritsch, Georg and Pickard, Chris J and
                  Salzmann, Christoph G and Michaelides, Angelos},
  journal={Phys. Rev. Lett.},
  volume={116},
  number={2},
  pages={025501},
  year={2016},
  publisher={APS}
}

@article{evans2009pair,
  title={Pair correlation function decay in models of simple fluids
                  that contain dispersion interactions},
  author={Evans, R and Henderson, JR},
  journal={J. Phys.: Condens. Matter},
  volume={21},
  number={47},
  pages={474220},
  year={2009}
}

@misc{li2025assessment,
      title={Assessment of First-Principles Methods in Modeling the Melting Properties of Water}, 
      author={Yifan Li and Bingjia Yang and Chunyi Zhang and Axel Gomez and Pinchen Xie and Yixiao Chen and Pablo M. Piaggi and Roberto Car},
      year={2025},
      eprint={2512.23940},
      archivePrefix={arXiv},
      primaryClass={physics.chem-ph},
      url={https://arxiv.org/abs/2512.23940}, 
}

\end{document}

% \title{Supporting Information: \emph{Ab initio} multiscale
% modeling of liquids made simple}
%\title{\tcb{Supporting Information: \emph{Ab initio} neural cDFT: A
%    unified framework for multiscale modeling of fluids}}

\title{Supporting Information: A unified machine-learning framework for \emph{ab initio} multiscale modeling of liquids}

\author{Anna T. Bui}
\affiliation{Yusuf Hamied Department of Chemistry, University of
  Cambridge, Lensfield Road, Cambridge, CB2 1EW, United Kingdom}
\affiliation{Department of Chemistry, Durham University, South Road,
  Durham, DH1 3LE, United Kingdom}

\author{Stephen J. Cox}
\email{stephen.j.cox@durham.ac.uk}
\affiliation{Department of Chemistry, Durham University, South Road,
  Durham, DH1 3LE, United Kingdom}

\date{\today}

\maketitle

%\tableofcontents
%\newpage

\section{Applicability to different MLIP architectures} 

\begin{figure*}[h]
  \includegraphics[width=\linewidth]{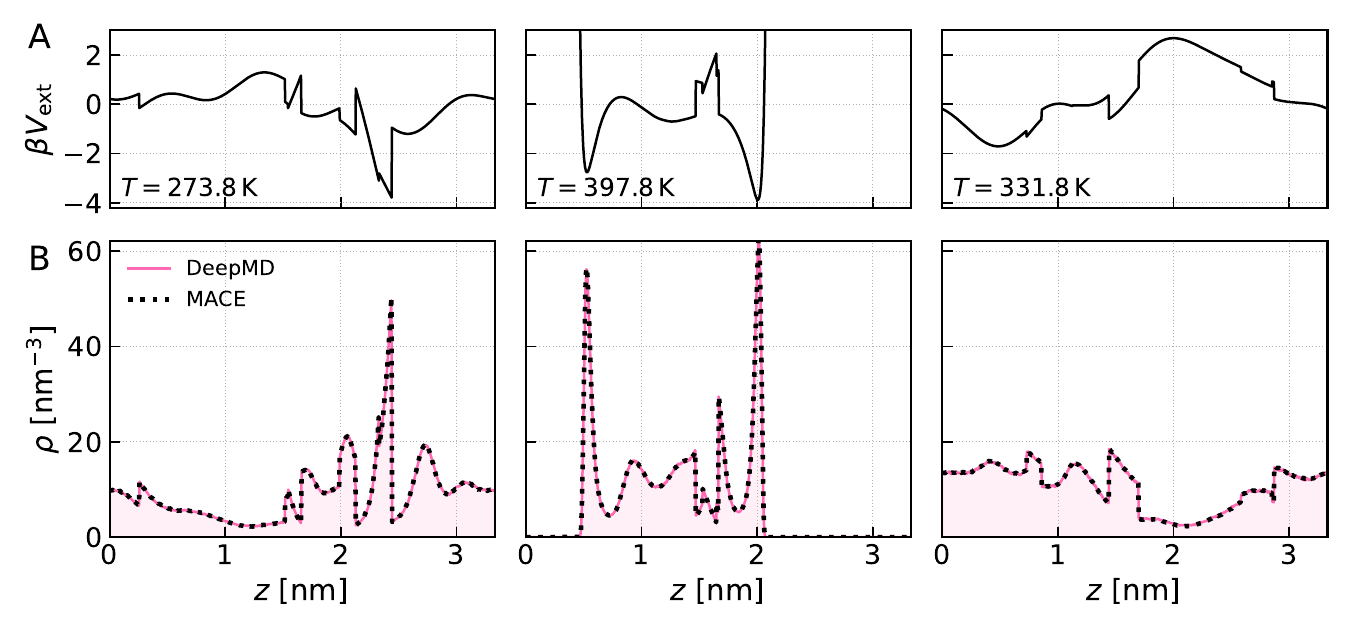}
  \caption{\textbf{Robustness to choice of MLIP architecture.}
    For random external potentials applied (A), the density profiles
    obtained from MD simulation (B) of PBE-D3 carbon dioxide are
      insensitive to the choice of DeepMD or MACE.}
\label{fig1}
\end{figure*}

To generate the training data for the neural cDFT, we have employed
MLIPs with the DeepMD \cite{Zhang2018} architecture for carbon dioxide
with all xc functionals considered (PBE-D3, BLYP, SCAN-rVV10). We also
used DeepMD for SCAN water. For RPBE-D3 water, we employed the HD-NNP
\cite{Behler2007} MLIP architecture. To gauge whether training with
the random $V_{\rm ext}$ is sensitive to the choice of MLIP
architecture, we also trained an MLIP employing MACE
\cite{Batatia2022} for PBE-D3 carbon dioxide, using the same dataset
from Ref.~\onlinecite{Mathur2023} as for the DeepMD MLIP. As seen in
Fig.~\ref{fig1}, results obtained with MACE are virtually
indistinguishable from those with DeepMD, suggesting that the
generation of training data for neural cDFT is robust to any
reasonable choice of MLIP architecture.

\newpage

\section{Multiscale prediction} 

\begin{figure*}[h]
  \includegraphics[width=0.95\linewidth]{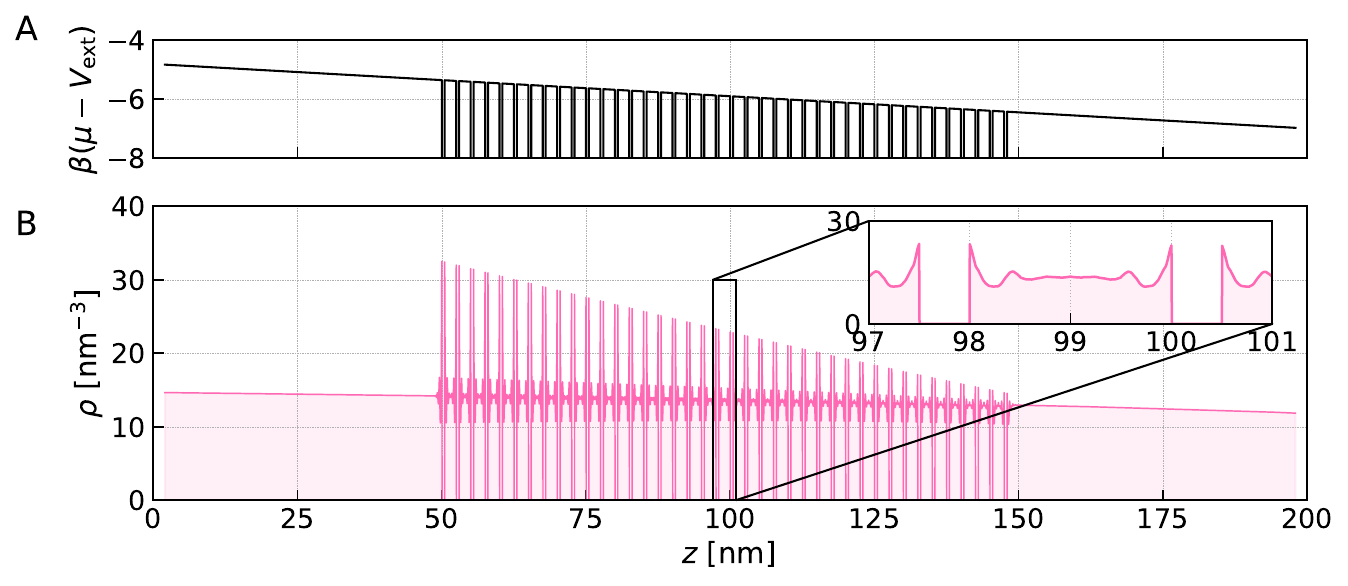}
  \caption{\textbf{Mesoscale prediction of PBE-D3 carbon dioxide.} For
    carbon dioxide at $T=300\,$K distributed across a multilayered
    membrane, with a concentration gradient modeled through a linear bias component in the
    external potential in A, the density profile is
    shown in B. }
\label{fig2}
\end{figure*}

The local nature of the trained neural functionals means that they are
able to make efficient predictions on mesoscopic length scales. As an
example, in Fig.~\ref{fig2} we present results for PBE-D3 carbon
dioxide in which $\beta V_{\rm ext}(z)$ decreases linearly over
200\,nm, along with 25 regions, each with a 1\,eV bias of thickness
0.5\,nm,
% are added
separated from each other by 2\,nm. It can clearly
be seen that \emph{ab initio} neural cDFT simultaneously describes the
large length scale redistribution of the fluid and the fine
microscopic structure. These results were obtained in approximately
one hour on a GPU.

\section{Additional results across interatomic potentials}

\begin{figure*}[h]
  \includegraphics[width=0.95\linewidth]{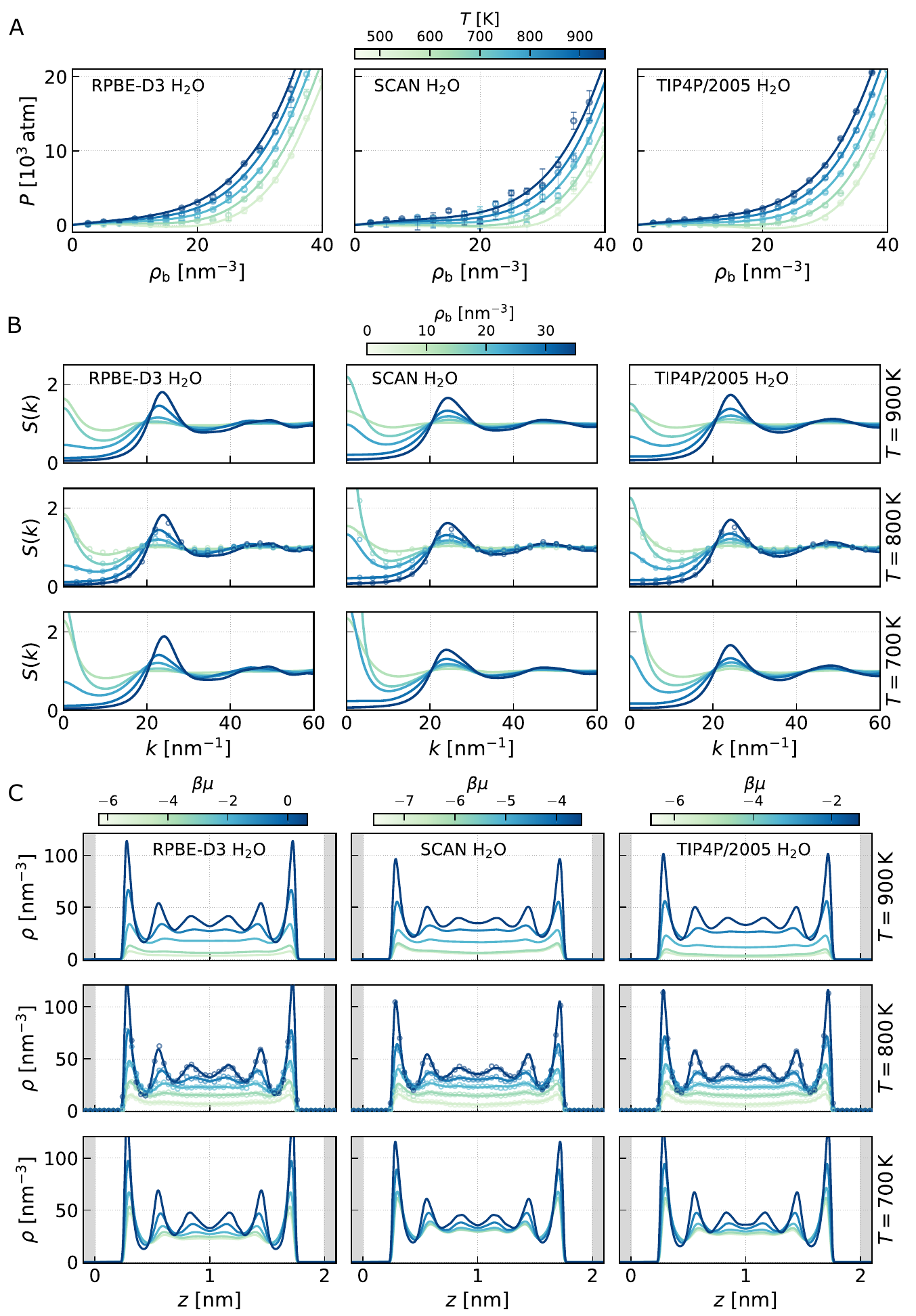}
  \caption{\textbf{\emph{Ab initio} structure and thermodynamics of
      water.} Neural cDFT prediction for water described with the xc
    functionals RPBE-D3 and SCAN, and the TIP4P/2005 classical force
    field are shown with solid lines. A: The equation of
      state. B: The structure factor. C: The structure of the fluid
      confined between two graphene sheets. Where atomistic
      simulations have also been performed, we show the corresponding
      results with symbols.}
\label{fig3}
\end{figure*}

\begin{figure*}[h]
  \includegraphics[width=\linewidth]{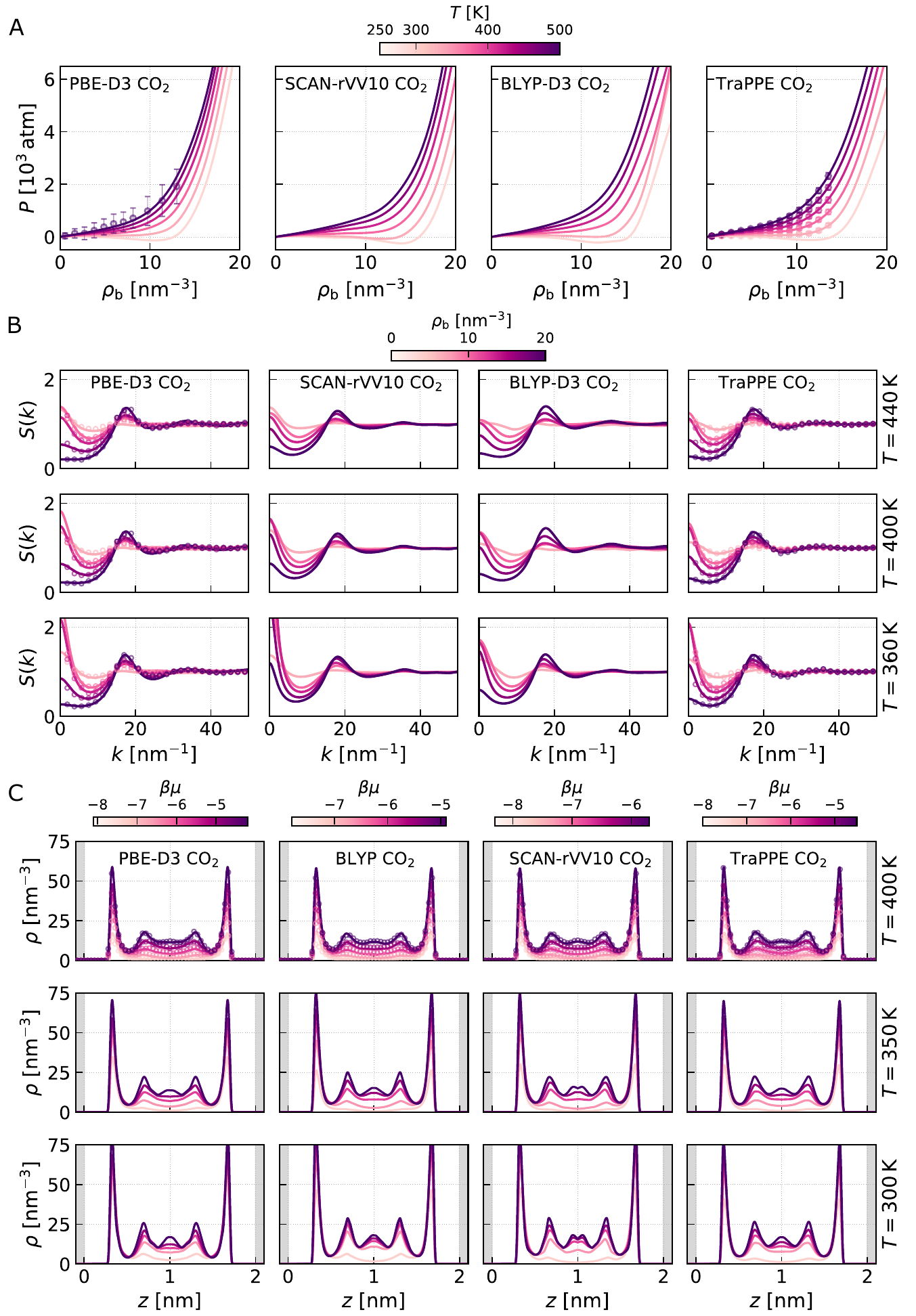}
  \caption{\textbf{\emph{Ab initio} structure and thermodynamics of
      carbon dioxide.} Neural cDFT prediction for carbon dioxide
    described with the xc functionals PBE-D3, SCAN-rVV10 and BLYP-D3,
    and the TraPPE classical force field are shown with solid lines.
    A: The equation of state. B: The structure factor. C: The
    structure of the fluid confined between two graphene sheets. Where
    atomistic simulations have also been performed, we show the
    corresponding results with symbols.}
\label{fig4}
\end{figure*}

Using neural cDFT, we show predictions across all interatomic
potentials investigated for the equation of state, the structure
factor, and fluid structure under confinement. Results for water are
shown in Fig.~\ref{fig3}, and those for carbon dioxide in
Fig.~\ref{fig4}. These correspond to genuine out-of-sample
  predictions, as equations of state and structure factors from MD
  simulations are not included in the training.

\newpage

\begin{figure*}[h]
  \includegraphics[width=0.8\linewidth]{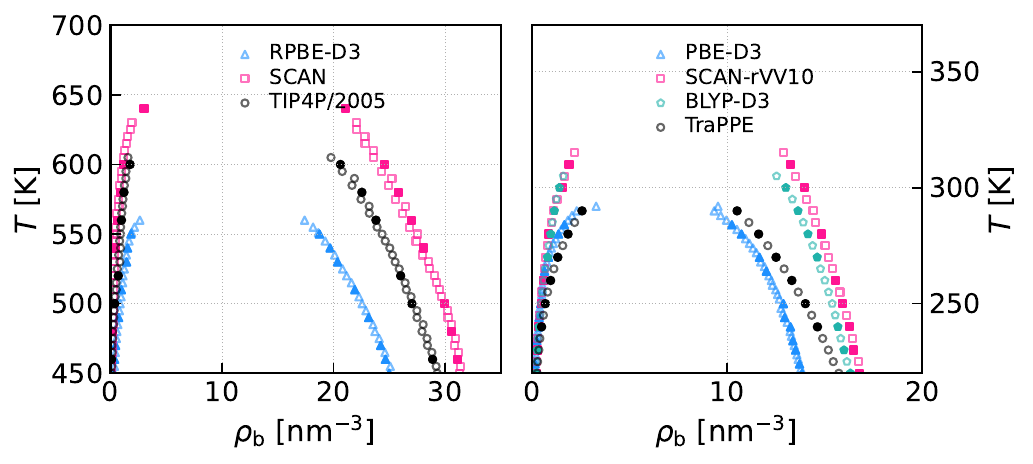}
  \caption{\textbf{cDFT prediction of liquid--vapor
        binodals}. Open markers indicate temperatures not included in
      the training set, while filled markers correspond to
      temperatures included in the training. These results demonstrate
      smooth interpolation across thermodynamic conditions.}
\label{fig5}
\end{figure*}

\begin{figure*}[h]
  \includegraphics[width=0.8\linewidth]{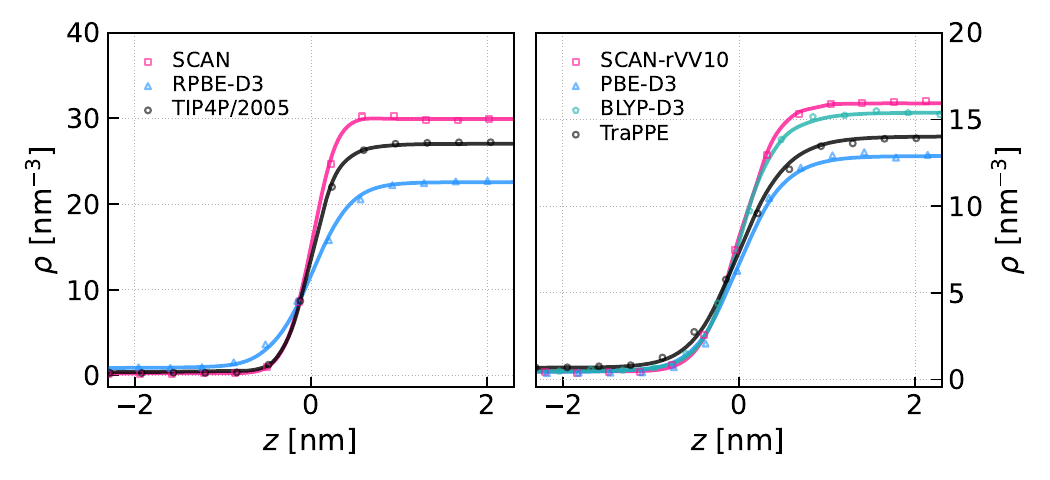}
  \caption{\textbf{The liquid-vapor interface.} Density profiles
      $\rho(z)$ at coexistence for water at $T=500\,$K (left) and
      carbon dioxide at $T=250\,$K (right), corresponding to Fig.~2B
      of the main text. Profiles are aligned by the Gibbs dividing
      surface, approximated by the point at which $\rho(z)$ is equal
      to half the density of the liquid phase.}
\label{fig6}
\end{figure*}

 The cDFT
predictions for liquid--vapor binodals are shown in
Fig.~\ref{fig5}, demonstrating \tcr{smooth interpolation of coexistence properties between discrete training temperatures.} Corresponding interfacial density profiles are
shown in Fig.~\ref{fig6}.

To quantify these results, Table~\ref{tab:errormetrics} reports
root-mean-square errors (RMSEs) between cDFT predictions and
simulation data.

\setlength{\tabcolsep}{7pt}
\begin{table}[H]
\centering
\begin{tabular}{c | c c c | c c c c}
\hline
\hline
 & \multicolumn{3}{c|}{Water} & \multicolumn{4}{c}{Carbon dioxide} \\
\cline{2-4} \cline{5-8}
RMSE & RPBE-D3 & SCAN & TIP4P/2005 & PBE-D3 & SCAN-rVV10 & BLYP-D3 & TraPPE \\
\hline
P [$10^3\,$atm] & 0.33 & 0.79 & 0.21 & 0.13 & - & - & 0.01 \\
$S(k)$ & 0.033 & 0.033 & 0.024 & 0.044 & - & - & 0.018\\
$\rho(z)$  [nm$^{-3}$] & 0.64 & 0.58 & 0.24 & 0.86 & 1.28 & 0.69 & 0.20\\
$\rho_{\mrm{v}}$  [nm$^{-3}$] & 0.33 & 0.26 & 0.16 & 0.15 & 0.17 & 0.04 & 0.05\\
$\rho_{\mrm{l}}$  [nm$^{-3}$] & 0.22 & 0.12 & 0.49 & 0.11 & 0.15 & 0.10 & 0.08\\
\hline
\hline
\end{tabular}
\caption{\textbf{Quantitative errors of cDFT predictions against simulations. Root-mean-square errors (RMSE) are reported for bulk pressure $P$, structure factor $S(k)$, density profiles $\rho(z)$, and coexistence densities $\{\rho_{\mrm{v}}, \rho_{\mrm{l}}\}$. For each observable,  the RMSE is computed over all overlapping data points for cDFT and simulations in the relevant variable (density, wavevector, spatial coordinate, or temperature).}}
\label{tab:errormetrics}
\end{table}

\begin{figure*}[h]
  \includegraphics[width=\linewidth]{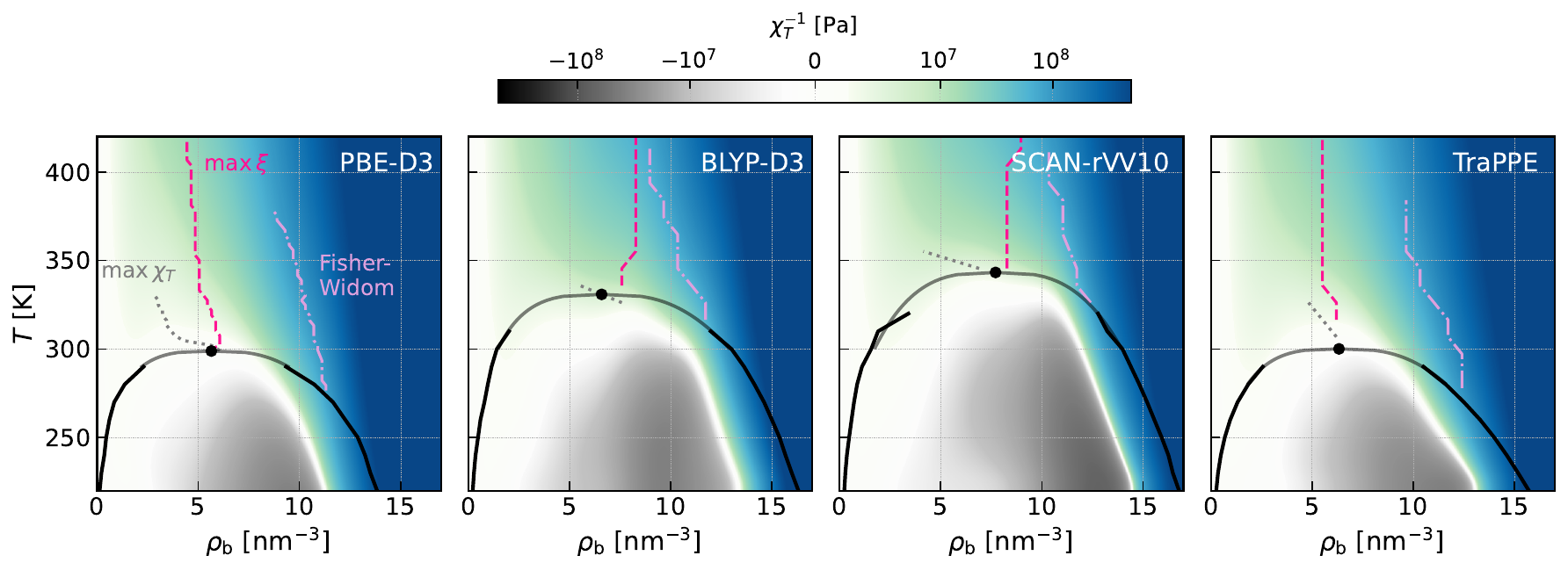}
  \caption{\textbf{Supercritical carbon dioxide across different
        interatomic potentials.} $\rho_{\rm b}$--$T$ phase diagram with
      $\chi_T^{-1}$ shown as a heat map. The Widom line obtained from
      $\max \chi_T$ is shown as a dotted line. The dashed line shows
      the Widom line from $\max \xi$, where $\xi$ is the correlation
      length. The dot-dashed line indicates the Fisher--Widom line,
      marking the crossover from monotonic to oscillatory decay of the
      total correlation function.}
\label{fig7}
\end{figure*}

In the main text, Fig.~4 shows results for PBE-D3 carbon dioxide;
  In Fig.~\ref{fig7} we present the corresponding results for the
  other interatomic potentials.  

\section{Thermodynamics of confined fluids}

In the main article, we consider single-component fluids confined
  between two graphene sheets separated by $H$. In the standard
  thermodynamic treatment of confined fluids
  \cite{HansenMcDonaldBook, Evans1987}, the exact differential of the
  grand potential is
\begin{equation}
  \mrm{d}\Omega  = -S \mrm{d}T - P\mrm{d}V - N\mrm{d}\mu +2 \gamma \mrm{d}A - \Pi A \mrm{d}H^\prime,
\end{equation}
where $S$ is the entropy, $\Pi$ is the disjoining pressure, $P$ is the
bulk pressure of the reservoir, and $\gamma$ is the substrate--fluid
interfacial tension. Note that $H^\prime$ is defined by a
  reasonable choice of dividing surface; in general $H^\prime \neq
  H$. As $V = AH^\prime$, it immediately follows that
\begin{equation}
  -\frac{1}{A}\left(\frac{\partial \Omega}{\partial H^\prime}\right)_{A, T, \mu } = P + \Pi,
\end{equation}
which we define as the effective pressure $\tilde{P}$. Even
  though in general $H^\prime \neq H$, $\tilde{P}$ is insensitive to
  the precise choice of $H^\prime$. This can be seen immediately by
  writing $H^\prime = H + \delta H$, such that
  $\mrm{d}H^\prime = \mrm{d}H$. For this reason, we adopt the simple
  choice $H^\prime = H$.

\section{Assessing thermodynamic consistency}

\begin{figure*}[h]
  \includegraphics[width=0.85\linewidth]{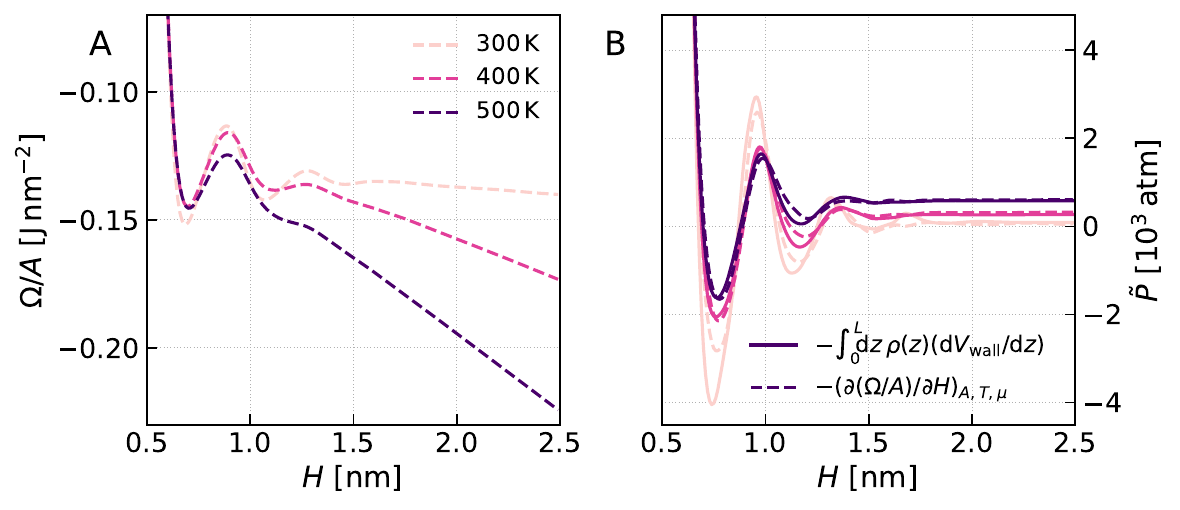}
  \caption{\textbf{Assessing thermodynamic consistency in neural
        cDFT.} A: $\Omega$ vs. $H$ for TraPPE carbon dioxide confined
      between two graphene sheets. B: Resulting $\tilde{P}$ compared
      to results in the main article, which were obtained by the
      structural route.}
\label{fig8}
\end{figure*}

As discussed in the main article, there are two routes to
  calculating $\tilde{P}$. The thermodynamic route is
\begin{equation}
      \tilde{P} = -\frac{1}{A}\left(\frac{\partial \Omega}{\partial H}\right)_{A, T, \mu },
\end{equation}
and the structural route is
\begin{equation}
  \tilde{P} = -\int^{L}_{0}\!\!\mrm{d}z\, \rho(z) \frac{\mrm{d} V_{\mrm{wall}}(z) }{\mrm{d}z}.
\end{equation}
That is, we can either calculate the derivative of the grand
  potential, or we can integrate
  $\rho(z)(\mrm{d}V_{\rm wall}/\mrm{d}z)$. In principle, both
  approaches should give the same result. In Fig.~\ref{fig8}A we show
  $\Omega$ vs. $H$ for TraPPE carbon dioxide confined between two
  graphene sheets. In Fig.~\ref{fig8}B, we compare the resulting
  $\tilde{P}$ to that presented in the main article, which were
  obtained by the structural route. At supercritical temperatures, we
  see that thermodynamic consistency is overall very good. At
  subcritical temperatures, more pronounced discrepancies are observed;
  this is due to an accumulation of numerical errors when integrating
through intermediate densities in the van der Waals loop
\cite{Sammuller2025}.

\section{Phase diagram}

\begin{figure*}[h]
  \includegraphics[width=0.45\linewidth]{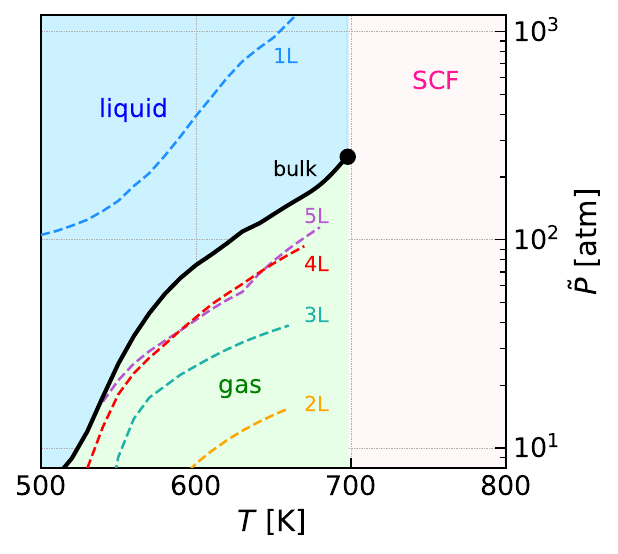}
  \caption{\textbf{Phase diagram of SCAN water upon confinement between graphene sheets.} Liquid--vapor phase diagram in the $\tilde{P}$--$T$ plane, for
  the different $H$ indicated.}
\label{fig9}
\end{figure*}

In addition to the $P$--$T$ phase diagram of confined water presented
in Fig.~3 in the main paper, where $P$ is the bulk pressure of the
reservoir, neural cDFT also allows construction of the $\tilde{P}$--$T$
phase diagram, where $\tilde{P}=P+\Pi$. We show this corresponding
phase diagram in Fig.~\ref{fig9}.

\newpage

\section{Hyper-DFT to obtain other equilibrium observables} 

While cDFT provides structural information based on the one-body
density $\rho$, hyper-DFT \cite{Sammuller2024hyper} is a recent
extension that states that any equilibrium observable of the
fluid can be written as a functional of the one-body density (the
  ``hyperdensity functional''). Here, as an example, we consider the
density of the hydrogen atoms $\rho_{\mrm{H}}$ in water as the
observable of interest; recall that we use the oxygen position to
  define the one-body density $\rho$. For an inhomogeneous fluid,
once the equilibrium density $\rho(z)$ is determined from solving the
Euler--Lagrange equation, the equilibrium hydrogen density can be
obtained by evaluating
\begin{equation}
  \rho_{\mrm{H}}(z) = \rho^{(1)}_{\mrm{H}}(z;[\rho],T),
\end{equation}
where $\rho^{(1)}_{\mrm{H}}(z;[\varrho],T)$ is the corresponding
hyperdensity functional.

\begin{figure*}[h]
  \includegraphics[width=0.8\linewidth]{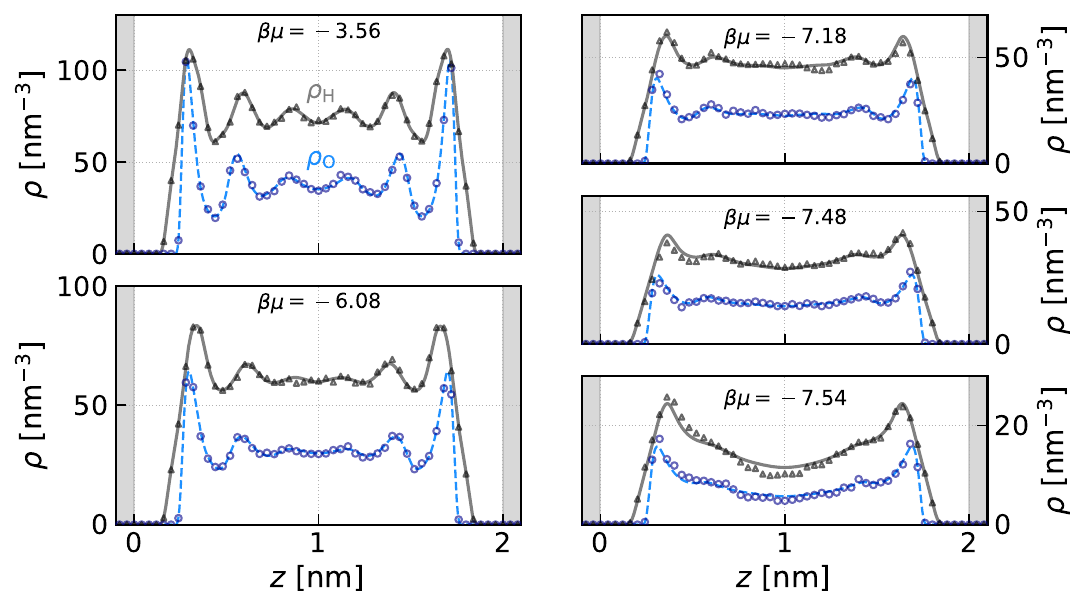}
  \caption{\textbf{Water's hydrogen atom density profiles
        obtained with hyper-DFT.} For SCAN water confined between two
    graphene sheets at different chemical potentials at
    $T=700\,$K, we show the oxygen atom density profiles
    $\rho_{\tcr{\mrm{O}}}=\rho$ obtained from neural cDFT (dashed blue line)
    and hydrogen atom density profiles $\rho_{\tcr{\mrm{H}}}$ obtained from
    neural hyper-DFT (solid gray line), in good agreement with MD
    simulation (symbols).}
\label{fig10}
\end{figure*}

To obtain a neural functional representation of
$\rho^{(1)}_{\mrm{H}}(z;[\rho],T)$, for each training simulation,
we sample not only the equilibrium one-body density $\rho(z)$ centered
on the oxygen atoms but also the equilibrium hydrogen density
$\rho_{\mrm{H}}(z)$. Then the mapping
$\{\rho(z),T\}\rightarrow \rho_{\mrm{H}}(z)$ is learned locally with a
neural network, analogous to the training of $c^{(1)}(z;[\rho],T)$
(see \emph{Methods} section in the main paper). The prediction of the
hydrogen density for confined water described with the SCAN xc
functional is shown in Fig.~\ref{fig10}.

%----------------------------------------------------------------------
\bibliography{references}
%----------------------------------------------------------------------